\documentstyle[12pt,aasms4,psfig]{article}
\def\ltsima{$\; \buildrel < \over \sim \;$}
\def\simlt{\lower.5ex\hbox{\ltsima}}
\def\gtsima{$\; \buildrel > \over \sim \;$}
\def\simgt{\lower.5ex\hbox{\gtsima}}
\def\kms{{\rm\,km\,s^{-1}}}
\def\kpc{{\rm\,kpc}}

\def\msun{{\rm\,M_\odot}}

\def\pc{{\rm\,pc}}

\def\s{\ifmmode \widetilde \else \~\fi}
\def\={\overline}

\def\spose#1{\hbox to 0pt{#1\hss}}

\def\etal{{\it et al.\ }}
\def\cf{{\it cf.\ }}
\def\eg{{e.g.,\ }}
\def\ie{{i.e.,\ }}
\def\lta{\mathrel{\spose{\lower 3pt\hbox{$\mathchar"218$}}
     \raise 2.0pt\hbox{$\mathchar"13C$}}}
\def\gta{\mathrel{\spose{\lower 3pt\hbox{$\mathchar"218$}}
     \raise 2.0pt\hbox{$\mathchar"13E$}}}
\def\Dt{\spose{\raise 1.5ex\hbox{\hskip3pt$\mathchar"201$}}}	
\def\dt{\spose{\raise 1.0ex\hbox{\hskip2pt$\mathchar"201$}}}	

\def\=={\equiv}

\def\dotsfill{\leaders\hbox to 1em{\hss.\hss}\hfill}

\def\des{Hatzidimitriou\ }
\def\Sgr{The Sagittarius dwarf spheroidal}
\def\sgr{the Sagittarius dwarf spheroidal}
\def\sgg{the Sagittarius dwarf}
\def\Sgg{The Sagittarius dwarf}
\def\com{center of mass}
\def\Gyr{{\rm\,Gyr}}

\def\msunpc{{\rm\,M_\odot/pc^3}}

\slugcomment{Submitted to the Astrophysical Journal}

\lefthead{Ibata \& Lewis}
\righthead{Galactic Indigestion}

\begin{document}

\title{Galactic Indigestion: \nl
Numerical Simulations of the Milky Way's Closest Neighbor}

\author{Rodrigo A. Ibata\altaffilmark{1,2} \& Geraint F. Lewis\altaffilmark{3,4}}

\altaffiltext{1}
{Department of Physics and Astronomy, University of British Columbia \nl
2219 Main Mall, Vancouver, B.C., V6T 1Z4, Canada}
\altaffiltext{2}
{Present address: European Southern Observatory \nl
Karl Schwarzschild Stra\ss e 2, D-85748 Garching bei M\"unchen, Germany \nl
Electronic mail: ribata@eso.org}

\altaffiltext{3}
{Department of Physics and Astronomy, University of Victoria \nl
PO Box 3055, Victoria, B.C. Canada V8W 3P7 \nl
Electronic mail: gfl@uvastro.phys.uvic.ca}

\altaffiltext{4}
{Department of Astronomy, University of Washington \nl
Seattle, Washington, U.S.A. \nl
Electronic mail: gfl@astro.washington.edu}



\begin{abstract}
Are  dwarf  spheroidal galaxies  dark matter  dominated?   We present N-body
simulations of  the   interaction between the  Milky  Way  and  its  closest
companion, \sgr\ galaxy, constrained by new  kinematic, distance and surface
density observations detailed in a companion paper.  It  is shown that there
is no possible self-consistent solution to the present existence of \sgg\ if
its distribution  of luminous matter traces   the underlying distribution of
mass.  The luminous component of the dwarf galaxy must therefore be shielded
within  a  small dark  matter halo.   Though   at present  we are  unable to
construct a fully  self-consistent model that includes  both the stellar and
dark matter components, it  is shown numerically that  it is possible that a
pure dark matter model, approximating the dark matter halo deduced for \sgg\
from analytical arguments, may indeed survive the Galactic tides.

The orbit of \sgg\  around the Milky Way is  considered, taking into account
the perturbative effects of the Magellanic Clouds.  It  is shown that at the
present time, the orbital period must be  short, $\sim 0.7\Gyr$; the initial
orbital  period for a  $10^9\msun$ model will have been  $\sim 1\Gyr$. It is
found  that a close encounter with  the  Magellanic Clouds may have occured,
though  the chances of  such an interaction  affecting the orbit of \sgg\ is
negligible.
\end{abstract}




%

\section{Introduction}

The Sagittarius  dwarf galaxy (\markcite{me94}Ibata,  Gilmore \& Irwin 1994,
\markcite{me95c}1995),  the  closest satellite  galaxy   of  the Milky  Way,
provides an  ideal laboratory  in which  the complex interactions  that take
place during  the merging  of galaxies may   be probed.  Motivated  by these
considerations, much information  has  now been  gained on its   kinematics,
metallicity and  stellar populations; the observational constraints obtained
hitherto   are  reviewed  in   \markcite{me97}Ibata \etal\ (1997;  hereafter
referred to as IWGIS).

A particularly interesting assertion that results from  an analysis of these
data is  that the  sheer existence  of  \sgg\ at the   present time is  very
surprising.  Accurate kinematic and distance data, which  now sample most of
the extent of \sgr, imply (subject to an assumption scrutinized in section~3
below) that this  dwarf galaxy has a  short orbital period around the  Milky
Way, less than $\sim  1$~Gyr.  Previously published numerical experiments of
the disruption of   this dwarf galaxy  (\markcite{vel95}Velasquez  \& White,
1995; \markcite{joh95}Johnston  \etal, 1995) showed  that it is  unlikely to
survive more than  a  few perigee  passages.  Taking  the results  of  these
simulations to  their logical  conclusion,  IWGIS argued that  the  observed
stellar population  cannot  trace the mass  of  that dwarf  galaxy, as \sgr\
would have been destroyed by the Galactic tides long ago.  A self-consistent
solution to the present existence of the dwarf can then only be found if the
requirement that  light     traces mass  is  relaxed.      Using  the simple
Jacobi-Roche tidal disruption criterion,  IWGIS proposed a solution in which
the stellar  component of the  dwarf galaxy is  enveloped in a halo  of dark
matter,  which  has a mass  profile such  that dark  matter  density  at the
photometric edge  of  the  dwarf  is   sufficiently  high to  impede   tidal
disruption.  To be consistent  with the observed  low velocity dispersion of
the stellar component  embedded therein,  the core radius  of  the dark halo
would have to extend out to the photometric edge of the system.

The dwarf spheroidal companions of the Milky Way have long been suspected to
contain large quantities of dense dark  matter \markcite{fab83}(\eg Faber \&
Lin 1983;  \markcite{irw95}Irwin \& \des 1995), so  the above conclusion for
the specific case  of \sgg\ is  perhaps not surprising; however, the density
profile of the  dark matter deduced  by IWGIS has important implications for
the nature  of the dark   halos and their constituents.   Most  of the dwarf
spheroidals contain  stars with a   broad range  of ages and  metallicities,
which  is   unexpected in  the  simplest  explanation   for their  low  mean
metallicities --- that chemical evolution was truncated by supernovae-driven
winds  (\eg  \markcite{san65}Sandage  1965; \markcite{dek86}Dekel  \&   Silk
1986); this  problem  may  be  alleviated with  the dark  matter  halo model
proposed by IWGIS, due an enhancement of the escape  velocity from the dwarf
galaxy.

These considerations about the dark matter content of \sgg\ have substantial
implications for the   currently-popular hierarchical clustering picture  of
structure formation, such as Cold-Dark-Matter  dominated cosmologies. A very
significant  accretion and merging   of  smaller systems  occurs during  the
evolution of a normal galaxy  like the Milky Way; is  this still an on-going
process?

In this  paper we aim to  examine  IWGIS' claims, redoing  their approximate
analytical   calculations  with numerical   disruption  experiments.   These
simulations will   be  constrained with  all   available relevant  data.  In
particular, we will   first investigate whether  the  assumption  adopted by
IWGIS in determining  the orbit of \sgg\  holds true,  and so establish  the
best-fit  orbit; and  secondly, we  will attempt  to  find a self-consistent
solution to  the present existence of  the dwarf  galaxy.  This provides the
necessary detailed analysis to determine the validity of IWGIS' assertion.

\section{Simulations}

The numerical  N-body simulations presented below   were performed using the
{\tt  box\_tree} tree  code  program (version 2.1),   kindly provided by  D.
Richardson (\markcite{ric93}Richardson 1993).    In all  simulations,    the
tree-code  opening angle was set  to $\theta=0.5$, and quadrupole correction
was used.

\subsection{Galactic potential model}

The forces of the Milky Way on the dwarf galaxy  are calculated by including
into the {\tt  box\_tree}   program one of   two  models  for  the  Galactic
potential.  For the majority of  the simulations, we  use an analytic model,
detailed in \markcite{joh95}Johnston {\it et al.}  (1995), which is composed
of a sum of three rigid potentials,  with the disk  component described by a
\markcite{miy75}Miyamoto-Nagai (1975) model,
\begin{equation}
\Phi_{disk} = - {G M_{disk} \over 
(R^2 + (a + \sqrt{z^2 + b^2})^2 )^{1/2}},
\end{equation}
the combined halo and bulge by a spherical Hernquist potential
\markcite{her90}(Hernquist 1990), 
\begin{equation}
\Phi_{sphere} = - {G M_{sphere} \over (r + c)},
\end{equation}
and the dark halo by a logarithmic potential,
\begin{equation}
\Phi_{halo} = {v^2}_{halo} \log (r^2 + d^2).
\end{equation}
In these expressions, $R$ and $z$ are in  cylindrical coordinates, while $r$
is  the radial distance in  spherical  coordinates; $M_{disk}  = 1.0  \times
10^{11} M_{\odot}$, $M_{sphere} = 3.4 \times 10^{10} M_{\odot}$, $v_{halo} =
128$~km/s, $a  = 6.5$, $b  = 0.26$, $c  = 0.7$ and $d  = 12.0$,  all in kpc.
This combination of parameters yields  an almost flat rotation curve between
$1$ and $30\kpc$.

Two of the  simulations carried  out below were  performed  with a potential
derived from  the  Galactic mass  model of \markcite{eva94}Evans  \&  Jijina
(1994). In this model, the disk component, described by a double exponential
disk, has radial   scale length $h_R=3.5  \kpc$  and a  Solar  neighborhood
surface  density of $\Sigma_0=48 \msun/\pc^2$.  We   further assume that the
vertical scale length of the disk  is $h_z=0.25 \kpc$,  and that the density
falls to zero at  $R = 5 h_R$.  The  potential corresponding to this density
distribution is  found by multipole expansion  of the Poisson equation using
an  algorithm  described in  \markcite{eng97}Englmaier  (1997).    The  halo
component is  described by  a  `power-law' halo, so   the potential  has the
following analytical expression:
$$
\Psi={ {v_0^2 R_c^\beta / \beta} \over 
{ ( R_c^2 + R^2 + z^2 q^{-2} )^{\beta/2} } }, {\hskip 1cm} \beta \ne 0,
$$
where $R$ and $z$  are   Galactocentric cylindrical coordinates, the    core
radius $R_c = 2\kpc$, $v_0 = 138 \kms$, and the exponent $\beta = -0.2$. The
parameter $q$ sets the oblateness of the  equipotential surfaces; two values
of this parameter were  probed,  $q = 1$  and  $q = 0.9$,  corresponding to,
respectively,  a spherical halo  and  a halo with  mass-flattening  of $\sim
0.7$.

\subsection{Dynamical friction}

As a massive object moves through the Galactic halo, the halo responds (in a
way that is dependent  on the density  and kinematics of  the halo,  and the
mass  and velocity of  the  object) to the  extra gravitational  attraction,
leaving  a wake of  halo material  behind the object.   This  wake, in turn,
exerts a force back on the object in the direction opposing its direction of
motion,   a phenomenon termed dynamical friction.    The effect of dynamical
friction    on the  galaxy  models  we  consider   is  significant,  but not
overwhelming.    This can be  seen by   considering  the orbital  decay time
$t_{fric}$  due to dynamical frictional  of a satellite  galaxy; an explicit
expression   is given by \markcite{bin87}Binney \&   Tremaine (1987) for the
particular  case of  a  satellite on  a  circular  orbit in  a potential  of
constant circular velocity.  The mean Galactocentric distance (averaged over
an   orbit) of \sgg\  is  $\simgt 25 \kpc$, while   the  most massive models
considered below have $M \sim 10^9 \msun$;  with these assumptions $t_{fric}
> 10 \Gyr$, so that such a model could have reduced its initial mean orbital
radius by at most a factor of $2$.   Since $t_{fric} \propto M^{-1}$, models
with $M \sim 10^8 \msun$ will hardly be affected at all.

The best way to account  for this frictional force is  to include a ``live''
halo of particles  in the simulation.   However, the implementation  of this
solution  is not  straightforward.  First,  for  the halo to respond  to the
passage of    the dwarf, sufficient spatial  resolution   is needed.   If we
require that there  be  at least $\sim   100$ halo particles in  the  volume
occupied  by  the  dwarf   at any   point  in its   orbit,  and   take  that
$\rho_{halo}(r) \propto r^{-2}$ (between the peri- and apoGalactic distances
of $15$ and $60 \kpc$), then $> 3 \times 10^6$ particles are needed to model
the halo (significantly more if one chose to populate the Galactic halo with
particles interior to $15 \kpc$).   Second, the distribution function of the
Galactic halo is currently unknown, so a range of  halo models would have to
be explored.  The computational   power required for these  calculations  is
well beyond that presently available to the authors.

So   to make this problem  tractable,   we take advantage   of an analytical
approximation; dynamical   friction  is included  into  the  {\tt box\_tree}
program by means  of an additional external  force on each of the simulation
particles, calculated as  follows.  First, we assume  that only the Galactic
dark halo  contributes any significant friction  to the simulation particles
(which is reasonable  given the peri-Galactic distance  of  the dwarf), that
the velocity distribution in  the dark halo is  a Maxwellian distribution of
dispersion $\sigma$ (of  value $220 \kms$,  set to be approximately equal to
the circular velocity of the halo), and that  the dark matter particles that
make up the Galactic dark  halo are of much lower  mass than the  simulation
particles  (which typically have  masses $> 10^4 \msun$).  These assumptions
simplify the  Chandrasekhar   dynamical friction formula to   the  following
relation (\markcite{bin87}Binney \& Tremaine 1987):
\begin{equation}
{ {d \vec{v}_i } \over {dt}} = - { {4 \pi \ln  \Lambda G^2 \rho M_a} \over {
{\overline{v}}^3  }} 
\biggl[  {\rm erf}(X) -  {  {2 X } \over  {\sqrt{\pi}}} e^{-X^2}
\biggl] \vec{\overline{v}}
\end{equation}
where $\vec{\overline{v}}$ is the mean velocity of particles within a radius
of $R_{a}$  of the $i$th   simulation particle, $M_a$ is  the  total mass of
particles      within     the     same     radius,    and       $X    \equiv
\overline{v}/(\sqrt{2}\sigma)$.  In all of  the simulations of the  internal
structure of \sgg, we set $R_a = 5 \kpc$, which is substantially larger than
the radius of bound particles in all  models.  This choice  of $R_a$ has the
effect  of essentially elliminating dynamical  friction on particles outside
the bound clumps.  The term $\ln \Lambda$ is  the Coulomb logarithm, defined
such that $\Lambda \equiv { b_{\rm  max} {V_0}^2 / G  ( M_a +  m ) }$, where
$b_{\rm  max}$ is the largest  impact  parameter between  the clump of  mass
$M_a$ and initial velocity $V_0$, and a galaxy with constituent particles of
mass  $m$.  Given that $M_a >>  m$, it follows that   $\Lambda \sim { b_{\rm
max} {\overline{v}}^2 /  G M_a }$.   Adopting $b_{\rm max}  =  100 \kpc$ (an
approximate upper  limit to  the  size of  the Milky  Way halo)  the Coulomb
logarithm (calculated  for  individual particles)  takes   on values in  the
(quite narrow) range $8 \simlt \ln \Lambda \simlt 12$ for the (rather large)
range  of  models   considered   below.   Clearly $\ln  \Lambda$  is   quite
insensitive to the guessed value of $b_{\rm max}$ adopted above.

There    is  a concern that the     Chandrasekhar relation overestimates the
dynamical friction, as  phase mixing in the   halo tends to dilute the  wake
behind the satellite.  However, this does  not pose a  problem to our study,
since we are primarily  interested in placing  lower limits on the  survival
time  of  \sgg.  The effect  of  overestimating  the frictional  force is to
increase the lifetime of the dwarf galaxy, as the  dwarf has to be placed on
an  initially longer period   orbit (hence less   disruption) than would  be
required in the absence of dynamical friction.

\section{Search for a self-consistent model: can light trace the mass?}

Our analysis is motivated by the claim that \sgr\ cannot have survived until
the present day if its light  traces its mass  given its observed extent and
velocity dispersion,  and  assuming that  the  orbit calculated by  IWGIS is
correct.  This last  assumption needs  to  be discussed  in detail.    IWGIS
fitted  the orbit  of  \sgr,  by considering   the  orbit of its  \com\  and
comparing the distance and projected line of sight  velocity along the locus
of  the orbit to mean  distance and mean  radial velocity measurements along
the    major axis  of   \sgg.   Successive    refinements   to the   initial
three-dimensional velocity of the \com\ allowed  iteration to a best-fitting
solution.  However, this calculation requires  two assumptions.  First, that
the observed direction of elongation of the dwarf  galaxy is closely aligned
with its motion vector. Numerical simulations  show that the tidal debris is
confined  to  the orbital  plane  of the progenitor (\markcite{oh95}Oh \etal
1995, \markcite{pia95}Piatek \&  Prior 1995, \markcite{joh95}Johnston  \etal
1995, \markcite{vel95}Velasquez \& White  1995). Since \sgg\ is  seen behind
the Galactic center, our line of sight must be  almost in the orbital plane,
so the observed   projected elongation should  indeed  be aligned  with  its
proper  motion vector to very good  approximation.  The second assumption is
that, for the purpose of  the orbit calculation, \sgg\ can  be modeled as  a
collection of massless  particles.  This proposition  is less clearly  true;
IWGIS argued that it is a fair approximation, but could not show it.

In  the remainder  of this  section,  we will consider   carefully a grid of
models that cover the    plausible parameter  space of  proto-dwarf   galaxy
structure.  We are aiming to answer definitively the question of whether any
combination of initial structural and kinematic parameters can guarantee the
survival of  \sgg\ until  the  present day,  while yielding  a   present-day
structure that is a good representation of  the observations.  We first list
the   constraints that    determine  the   allowable  orbits, the    initial
distribution function, and the allowable evolution of our models.

The following parameters determine the orbit:

(i) The distance to the \com\  is $25\pm 1 \kpc$ (\markcite{me94}Ibata \etal
1994, \markcite{mat95a}Mateo 1995).

(ii) The  radial  velocity of the  \com\  in a non-rotating  reference frame
centered at the present position of the Sun is $171 \pm 1 \kms$ (IWGIS).

(iii) As discussed above, the proper motion  vector of \sgg\ must be aligned
parallel to the Galactic  coordinate line $l=5^\circ$  (\markcite{me94}Ibata
\etal 1994, IWGIS). This puts it on essentially a polar orbit. The component
of the   proper motion of   the central regions  of  \sgg\ in  the direction
perpendicular to the Galactic plane has been measured (\markcite{irw96}Irwin
\etal   1996, IWGIS),   indicating that  it   is moving   northwards  with a
transverse velocity of $250 \pm 90 \kms$.

The following parameters determine the initial distribution function:

(iv)  A firm  lower bound to  the present  total  mass is   $10^7 \msun$, an
estimate  based on starcounts    of red giant  members (\markcite{me94}Ibata
\etal\ 1994).  In particular, that early  work covered only a small fraction
of the extent of the galaxy near its photometric center. (Note that the most
massive globular cluster of \sgg\ system, M54, alone has a mass of $2 \times
10^6 \msun$ ---  \markcite{pry93}Pryor  \& Meylan  1993).  However,  a lower
limit  of  $\sim  10^8  \msun$ is  obtained   from comparison   of  its mean
metallicity     to trends   observed     in  the  dwarf   galaxy  population
(\markcite{me95}Ibata \etal\ 1995); since this last estimate better reflects
the {\sl  initial} total mass (rather than  the total mass after significant
dynamical evolution in the  tidal field of  the Milky  Way), we adopt  $\sim
10^8 \msun$ as the lower limit of the initial mass of the models.

(v) The present concentration of the stellar  population in \sgg\ is $c \sim
0.5$ (IWGIS).  The initial stellar  concentration, is of course unknown, but
may be estimated by comparison to other ellipsoidal  galaxies.  Of the Milky
Way      dSphs, the   most    concentrated    is   Sculptor, with   $c=1.12$
(\markcite{irw95}Irwin  \& \des  1995).   As a firm   upper limit we adopt a
maximum  initial concentration of $c=2.1$ in  our models. (The value $c=2.1$
corresponds to a  central value of  $(\Psi/\sigma^2)_0 = 9$ in the formalism
of \markcite{bin87}Binney \&  Tremaine 1987, which   we follow below,  where
$\Psi$ is the relative potential and $\sigma$  is a model parameter related,
but  not equivalent, to the   central velocity dispersion).   This choice is
conservative,   since only dynamically  very   evolved stellar systems  have
significantly higher concentration.

(vi) It was observed that the models never evolved  to a state with a higher
central velocity dispersion; we therefore  investigated models with velocity
dispersion equal to, or higher than the observed present velocity dispersion
of the dwarf.

While the conditions that determine acceptable evolution are:

(vii) At  the end of  the integration, required to be  the  present day, the
\com\ of the model must be located at the observed position of the center of
\sgg: $R=25 \pm 1  \kpc$, $(\ell=5^\circ, b=-14.5^\circ$). The dispersion of
radial velocities  at the center of  the  model must be  consistent with the
observed value,    $\sigma_v=11.4 \pm  0.7  \kms$ (IWGIS).     The projected
velocity and distance gradients must also match the observations.

(viii) The final minor axis  half-mass radius must  match the observed minor
axis half-brightness radius $R_{HB} \sim 0.55 \kpc$ (IWGIS).
\footnote{The half-brightness radius was determined from the distribution of
red clump stars in \sgg. In the absence of a dark matter component, $R_{HB}$
will be equal to the half-mass radius only if there  has been no significant
mass segregation.}

(ix)  \Sgr\ cannot have lost most  of its luminous   mass.  It contains four
globular clusters and  a  substantial stellar component concentrated  over a
region less than $30^\circ$ in length. For  instance if \sgg, as observed at
the present time, had  only 25\%  of its  original mass within  the observed
extent (\ie within  $5 \kpc$ of  the photometric center), the remaining 75\%
--- in  this case  $\sim 16$ globular   clusters and a  substantial  stellar
population --- would  be visible  as a dense  ring of  globular clusters and
tidally disrupted debris   around the  Milky  Way.  This, however,  is   not
observed.  We therefore stipulate that, at the end point of the integration,
an acceptable model  should  retain $25  \%$ of  the initial mass  within $5
\kpc$ of the photometric center.

(x) \Sgr\ must retain  a central concentration for at  least $12 \Gyr$  (the
age of its dominant stellar population \markcite{fah96}Fahlman \etal 1996).

To start  the N-body simulations, a suitable   initial position and velocity
for the center of mass of the model must be found. To this end, a point-mass
particle is first integrated back in time  for 12~Gyr in the chosen Galactic
potential.  The information needed for  this calculation is: the present 3-D
position of  the \com, its radial velocity,  and the  projected direction of
the velocity vector of the \com, all of which are accurately known; plus the
proper  motion, which is  poorly  constrained.  Specific  to each model,  to
calculate the orbital decay due to dynamical friction, one also requires the
model mass,  and the mass  loss rate due to the  tidal disruption. Since the
mass loss rate is  not known before the  simulation  has been completed,  we
assume  that no mass loss occurs;  this strategy increases the survivability
of the  models (since the  model has an  initially longer orbit), consistent
with our aim of  placing a lower  limit on the survival  time of  \sgg.  The
large  inaccuracy of the proper  motion measurement means that the magnitude
of  the  present  velocity of \com\   of  the dwarf,  $|v_{com}|$, is poorly
constrained; to circumvent this    problem  we consider four  orbits    with
$|v_{com}|=332 \kms, 362 \kms, 392 \kms$ and $422 \kms$, which we label `a',
`b', `c'  and `d' respectively.   Orbit `a' is the   best-fit orbit found by
IWGIS.  It is uninteresting to simulate models where $|v_{com}| < 332 \kms$,
as this  leads  to  shorter periods  than  that  of orbit `a'   which simply
aggravate  the   survivability problem.  The  $R$--$z$  structures  of these
orbits calculated  for a test particle  of negligible mass (\ie no dynamical
friction) are shown in Figure~1, where we have integrated the orbits back in
time  for  12~Gyr  under the   influence  of  the fixed  Galactic  potential
described in \markcite{joh95}Johnston {\it et al.}  (1995).

The initial  configuration  of positions  and  velocities is constructed  by
choosing   particles   randomly from  a   King   model (\markcite{kin66}King
1966). King models fit the surface brightness of dSph quite acceptably (\eg\
\markcite{irw95}Irwin \& \des 1995), and  are therefore a reasonable  choice
for the  distribution  function  of the protogalaxy.   Three  parameters are
required  to describe the  model: following \markcite{bin}Binney \& Tremaine
(1987), these are the ratio of  the central value  of the relative potential
$\Psi$ to the square of the $\sigma$ parameter, \ie $(\Psi/\sigma^2)_0$, the
parameter $\sigma$, and the central value of the mass density $\rho_0$.

The starting  parameters  of  the King  models  are given  in  Table~1.  The
columns of this table  list: (1) the  model identification label, where  the
letter  following the hyphen   gives  the orbit  label;   (2) the value   of
$(\Psi/\sigma^2)_0$;    (3) the central   mass   density, $\rho_0$; (4)  the
$\sigma$ parameter of the model; (5) the one-dimensional velocity dispersion
of the model, $\sqrt{\overline{{v_x}^2}}$;  (6) the  half-brightness radius,
$R_{HB}$; (7) the tidal radius, $R_T$; (8) the total mass of the model, $M$;
and (9) the number $N$ of equal mass particles in the simulation.

Each  structural model (listed in  Table~1)   was evolved in simulations  of
either 4000 or 8000 particles.  To a reader  familiar with the present state
of numerical  experiments  in   cosmology,  simulations of  a  few  thousand
particles may   appear ridiculously   inadequate.  However,  the   time-step
choosing routine in \markcite{ric93}Richardson's (1993)  code, adapts to the
very   short crossing time   of particles the in   the  dwarf galaxy, and to
changing forces on the dwarf due to the external potential,  which vary on a
short timescale due to  its rapid orbital  velocity.  The limitation of 8000
particles,  matching previous  work (\markcite{joh95}Johnston  \etal\  1995,
\markcite{vel95}Velasquez  \& White  1995,  \markcite{pia95}Piatek \&  Pryor
1995),  was chosen because a  single such  simulation requires approximately
one month of Sparc Ultra II cpu time to complete.

A significant concern with N-body  simulations is that  the behavior of  the
modeled system may  not reflect the true  dynamics, since due to the present
computational limitations,  each simulation particle usually represents many
stars  (or  dark matter   particles).   Thus  in the   simulations,  violent
interactions of close particles are  more destructive than in reality, while
the diffusion of particles up to the escape velocity of the system, known as
``evaporation'',  leads to    faster  dissolution.   The effect of     close
encounters in the simulations   may  be reduced by  artificially   softening
gravity forces; we  chose a  smoothing  length  of $100  \pc$, matching  the
simulations  of \markcite{vel95}Velasquez   \& White  (1995).   However, the
effect of evaporation, being a  result of long-range interactions, cannot be
avoided.   The evaporation time,   $T_{ev}$, of a   King model  is typically
between    1\%  and   6\%   of the   half-mass    relaxation time,  $T_{rh}$
(\markcite{joh93}Johnstone 1993).    With equal   mass particles,   we  find
(\markcite{spi87}Spitzer 1987):
$$T_{rh}= \Bigg({{2}\over{3}}\Bigg)^{1/2} 
{ { ({\overline {v^2}})^{3/2} N} \over
{ (M^2 / 2 V_h) 4 \pi G^2 \ln \Lambda } },$$
where $\overline {v^2}$ is the mean-square velocity of the particles, $N$ is
the  number of simulated particles,  $M$ is the  mass of the model, $V_h$ is
its half-mass  volume,  and $\ln \Lambda$  is  the Coulomb  logarithm.   The
evaporation timescale of our models  (listed in column  10 of Table~1, using
Figure~2  of  Johnstone  1993  to  relate $T_{ev}/T_{rh}$  as a  function of
concentration)  is   longer than  the  $12  \Gyr$   integration time of  the
simulations, except  in the   case of  some  concentrated models  with  high
central density (K11 --  K15).  Evaporation effects  are therefore likely to
be significant for the models  K11 to K15, so for   these models the  quoted
survival times are lower limits.

For comparison of    the disruption rates  between   models, we define   two
timescales:  the time of ``complete   disruption'', $T_{dis}$, defined to be
that time at which  90\% of the  simulation particles lie outside  the Roche
radius; and $T_{ok}$, the time at which the  model becomes too unbound to be
consistent with observations, which we set  (for reasons mentioned above) to
be the time at which more than $25\%$ of the initial  mass lies further than
$5 \kpc$ from the central concentration.  The Roche radius  of the system is
calculated  by determining  the  radius at  which   the mean density  of the
simulation particles becomes  less   than twice  the Galactic mean   density
interior to the previous periGalactic distance of the orbit (recall that the
periGalactic  distance  decays  due to  dynamical  friction). Clearly, these
calculations require the center of  the model to be  known.  To this end, at
every  output timestep, we  find the simulation  particle from which the rms
deviation to its nearest 100 neighbors is smallest, and choose this particle
to represent the center of the dwarf galaxy (or  dwarf galaxy fragment).  In
Table~1, the quantities $T_{dis}$  and $T_{ok}$ are listed, respectively, in
columns (11) and (12); however, when $T_{dis}$ or $T_{ok}$ exceed $12 \Gyr$,
the same entries list, respectively, the fraction  (in percent) of particles
within the Roche  radius, and the fraction of  particles within $5  \kpc$ of
the center of the model.

To compare the velocity dispersion of  the simulated models to observations,
we calculate the radial velocity  dispersion of particles within a projected
radius of $3^\circ$ from  the center of the  models, at the simulation  time
$T_{ok}$,  as  seen  from the    position    of the  Sun.   This   quantity,
$\Big(\sqrt{\overline{{v_r}^2}}\Big)_{T_{ok}}$, is listed  in column (13) of
Table~1. Finally, the  minor axis half-brightness  radius, calculated at the
simulation   time $T_{ok}$, from particles  in  a $100\pc$  strip across the
minor axis of the model, is listed in column (14).

\subsection{The King model simulations}

We started our simulations with the  model K1-a, whose initial configuration
fits best the observations of the present state of  \sgg. In particular, the
one  dimensional velocity  dispersion  $(\overline{{v_x}^2})^{1/2}$ was  set
close to the observed value of $11.4 \kms$,  the minor axis half-mass radius
was set close to the observed value of $0.55 \kpc$, and the model was placed
on   the  best-fit orbit  found  by  IWGIS  (\ie  similar   to the  orbit of
Figure~1a).  The model disrupted rapidly,  however; with complete disruption
occurring  after only $5.3  \Gyr$.   These models confirm, using constraints
from a more complete data  set, the findings of \markcite{vel95}Velasquez \&
White (1995) and \markcite{joh95}Johnston \etal  (1995) that models of \sgg,
in which light traces mass,  are fragile.  If so, one  may deduce that \sgg\
will be completely disrupted within $\sim 5 \Gyr$, providing a source of new
stars  and globular  clusters to the   Galactic  halo.  At  the  end  of the
integration, at $12 \Gyr$,  the structure of the  galaxy remnant is shown in
Figure~2.  No bound concentration of particles  remains; instead, streams of
particles populate the orbital path of the former dwarf galaxy.

As  detailed  above,  the orbit  presented  by  IWGIS  was calculated  using
massless particles to trace the motion of \sgg.  Under that assumption, they
found  that  the observed radial velocity  gradient   implied a short period
orbit like  that shown in Figure~1a.   However, it is worth checking whether
the internal self-gravity of the dwarf galaxy could  act to accelerate stars
along  the   tidally distorted  major axis in   a way  that would  mimic the
velocity gradient of a shorter period orbit.  We therefore performed further
simulations   (K1-b, K1-c, K1-d), where   the model was placed progressively
longer period orbits  similar to those shown  in Figure~1 on panels `b', `c'
and `d', respectively. The model K1-b disrupted completely before the end of
the integration, while the models K1-c and K1-d retained an intact core (see
Figures~3 and~4 and Table~1).  However, the  major axis velocity gradient at
the end-point of   K1-c and K1-d is  much  larger than that  allowed  by the
observations, as we show  in Figure~5.  Furthermore, the velocity dispersion
at the simulation end-point is too low, inconsistent with the observations.

The initial structure  of the K1 models gives  a good  representation of the
present  state of   \sgg.  Since these    fail to  survive in the   Galactic
potential, one can immediately deduce that either there has been significant
structural evolution  from the initial state,  or that light does  not trace
mass in \sgg. In the remainder of this section we explore  the first of these
options, delaying discussion of the second option to section~4.

The King models K1 --  K8 were simulated to  investigate the plausible range
of initial models  with  fixed  $(\Psi/\sigma^2)_0=2.0$ (equivalent   to  an
observed  concentration  $c \sim 0.7$) which  are  consistent with the setup
constraints  listed above.  The central density  is increased from K1 to K8.
Higher central density   leads to smaller  half-mass and  tidal radii, which
should yield more robust models.  Indeed, the sequence of  models K1, K3 and
K5, which  were  set up  to have initial  velocity dispersion   equal to the
observed  present  velocity dispersion   of the  dwarf,  are  a  sequence of
increasing   robustness.  The model  K5-a   is  sufficiently  robust to   be
consistent  with the survival constraint,  it has a radial velocity gradient
also in  good   agreement with  observations;   however, its   final central
velocity  dispersion,  $\sigma_v=9.4   \kms$, is  approximately   3 standard
deviations lower than the observed value of $11.4 \pm 0.7 \kms$.

Since the  central  velocity dispersion  of our  models was found  to always
decay as the disruption proceeds,  we also simulated the  K2, K4, K6, K7 and
K8   models, where the  initial  value of the   velocity  dispersion was set
substantially  higher than  the observed present  day  value.  The K2 and K4
models,  like  their  lower   velocity dispersion counterparts   K1  and K3,
disrupted rapidly.   Similarly, the very  high  velocity dispersion model K7
was also quickly destroyed by the Galactic tides.

However, the K6-a  and K8-a models match  the velocity dispersion and radial
velocity   observations   quite  well,   and    satisfy  the  survival  time
constraint. Contour maps of these models at the end point of the simulation,
are shown in Figure~6. The axis ratios of the remnants are in good agreement
with the  deduced 3:1:1  ratio  (IWGIS).   However,  the final   minor  axis
half-brightness   radii  of  both    these models   is  $R_{HB}=0.15  \kpc$,
inconsistent with the observed value of $R_{HB} \sim 0.55 \kpc$.

More concentrated models may survive longer if their inner regions manage to
stay  stable while diffuse outer layers  are lost  to  Galactic tides.  This
notion led us  to undertake simulations   with the models  K9 --  K15, whose
concentration  $c=1.25$  ($(\Psi/\sigma^2)_0=6.0$) is  slightly  higher than
that of the  Sculptor dSph (the most  centrally concentrated of the Galactic
dSph), which has $c=1.12$ (\markcite{irw95}Irwin  \& \des 1995).  From K9 to
K16 we  increase the central density  while varying the velocity dispersion,
thereby sampling  a large range  of plausible models  at this concentration.
It   was found that models   on  the `b',  `c'  and  `d' orbits always  have
projected major  axis kinematics  that  are inconsistent  with observations.
Surprisingly,  the model K13-a survives the  interaction  with the Milky Way
almost  completely   intact,   despite  its  relatively   short  evaporation
timescale; this suggests that the method of \markcite{joh93}Johnstone (1993)
overestimates the evaporation rate in some situations.

Of  the models  with  $(\Psi/\sigma^2)_0=6.0$, K10-a managed  to survive the
Galactic  tides,  and  give    a   good representation  of   the   kinematic
observations.   Again, however, the final  minor axis half-brightness radius
is $R_{HB}=0.14 \kpc$, inconsistent with the observed value.

To check the survivability of less concentrated models we  ran models K16 --
K20.    As found above, only  those   models on the  `a'   orbit are able to
reproduce the observed radial velocity  gradient.  Of these models, K17-a is
sufficiently  robust and gives   a   good  representation of  the   observed
kinematics. Yet  its final  half-brightness  radius, $R_{HB}=0.17  \kpc$, is
again inconsistent with the observed value.

\subsection{The effect of initial major axis rotation}

Could \sgg\ be a  tumbling body, rotating along its  major axis, so that the
observed elongation is  a consequence of  rotation?  This seems unlikely, as
observations of   dwarf  ellipticals, in  particular,  NGC~147,  NGC~185 and
NGC~205 (\markcite{ben90}Bender  \&  Nieto  1990),  show that  they  are not
rotationally flattened. Nevertheless, could tumbling  either imply a  longer
period orbit or lead to less disruption?  IWGIS showed that clockwise linear
rotation in  the sense of Figure~1 implies  a  shorter orbital period, while
longer periods are  deduced for anti-clockwise  rotation.   By comparison to
the ellipsoidal stellar models of \markcite{bin78}Binney (1978), they argued
that given the observed axis ratios (3:1:1), the maximum rotational velocity
is likely to be   at most equal to   the velocity dispersion, \ie $\sim   11
\kms$.  Taking the  then outermost major axis  field (at $b=-24.5^\circ$) as
the major axis  limit of the system,  they deduced orbital periods  of $\sim
0.6  \Gyr$  and   $\sim    1.6  \Gyr$  for,   respectively,  clockwise   and
anti-clockwise linear rotation of the  major axis  around  the \com.   Since
clockwise  rotation  leads to  shorter orbital  periods  and  so  to shorter
survival times, only anti-clockwise rotation need be considered here.

It is interesting  to note a general result  valid for all of the spherical,
initially non-rotating,  models  we  investigated.   As the   dwarf  becomes
affected by  the Galactic tides, its shape  becomes elongated such  that its
leading edge is closer  to the Galactic center  than its trailing edge.  The
resulting  force  gradient torques the   now  slightly bar-like dwarf, which
progressively gains spin angular  momentum (presumably at  the expense of  a
small  amount of  orbital  angular momentum),  until its outer  stars become
centrifugally unbound.  The sense of the spin provided by  the Galaxy to the
dwarf is always the same as  the sense of the orbit  of the dwarf around the
Galaxy. Placed  on the  same orbit,  it  transpired  that models  which have
initial rotation in  the  same sense as   their orbit dissolve  faster  than
models whose initial rotation is in the opposing  sense, apparently due to a
cooperation with  the  Galaxy in the   centrifugal  expulsion of  the  outer
particles.  Since the proper motion measurement  indicates that the orbit of
\sgg\ is anti-clockwise in the sense of  Figure~1, anti-clockwise major axis
rotation is expected to promote disruption.

The bar-like distortion  induced  by the tides  near periGalacticon  to  the
non-rotating  models does not   significantly affect the  symmetry of  their
central  regions (say defined as that  region whose radius contains half the
mass  interior  to the  Roche  limit), which  remain approximately spherical
until  just before complete  disruption.    Thus the structure of  initially
non-rotating  models is   very different  to   that of  models  that have  a
significant fraction of their kinematic energy in rotation; in isolation the
latter  become  prolate triaxial bars   (see \eg\  \markcite{bin87}Binney \&
Tremaine 1987).  On orbiting around  the Galaxy, a prolate  bar with a major
axis  rotation vector  parallel  to the rotation   vector of its  orbit will
periodically point its major axis directly at the Galactic center.  With the
above axis  ratios (3:1:1),   the   dominant tidal   force  along the   line
connecting the  center of  the  Milky Way  and  \sgg\ will   be periodically
approximately 3 times  larger than in  a similarly-shaped non-rotating model
(the latter, as  discussed  above, align themselves approximately  along the
orbit).  Thus tumbling will lead to  greater tidal disruption, worsening the
survivability problem.

\subsection{Summary: Constant $M/L$ models}

In conclusion, we have attempted to model the interaction  of \sgg\ with the
Milky Way under the  assumption that the  observed  population of red  clump
stars traces the underlying mass distribution. With a few exceptions, these
models had $M/L \simlt 10$.

All the  models  placed initially  on  the `a' orbit give  rise  to a radial
velocity gradient  that  is in good agreement   with the observations, while
those models on the longer period `b', `c' and `d' orbits give progressively
worse fits to the radial velocity data.  This gives a strong indication that
the true orbit of \sgg\ is close to the `a' orbit,  which has a short period
$T \simlt 1 \Gyr$, and implies that there have been many collisions with the
Galactic disk in the past.

A thorough search of parameter space with  a grid of  20 models has revealed
several models (K5-a, K6-a,   K8-a, K10-a, K13-a  and  K17-a) that  retain a
bound core at the    end of the simulation,   and  fit the  radial  velocity
profile.  Of these, models K6-a, K8-a, K10-a and K17-a also fit the observed
radial velocity dispersion.   However,  no model manages  to  survive and be
consistent with the observed minor axis width of \sgg:  all models that have
initially large half mass radii are rapidly destroyed by the Galactic tides.

In all  the models,   the half  mass radius   becomes smaller as  disruption
proceeds. So large  initial half mass radii are  required to match the final
$R_{HB} =  550\pc$. Yet any such model  is rapidly destroyed by the Galactic
tides.

It must be concluded that at least one of the  assumptions laid out above is
not valid.   IWGIS  argued that the   most  plausible  explanation for  this
quandary  is that the  visible component of \sgg\  resides  within a halo of
dark matter  which protects it  against  Galactic tidal  forces.  We further
explore this claim below.

\section{A heavy course of dark matter}

Having  shown that  models  with constant  $M/L$  are unacceptable,  we next
investigate the survivability   of models with  dark matter  halos.

IWGIS used  the  analytical  Jacobi-Roche  criterion  to  show  that   it is
plausible that the stellar  component of \sgg\  could easily resist Galactic
tides if it was embedded in  a dark matter  halo.  For simplicity, the model
presented  was  a homogeneous sphere of   density $\rho=0.03 \msunpc$, which
extended out  to a cutoff  at  the visible  minor axis  limit of the system,
$r_c=1.5   \kpc$.  \markcite{bin87}Binney  \&   Tremaine (1987) discuss  the
derivation of  distribution functions   for  spherical systems   from  their
density profile.   They show that the  distribution  function of a spherical
stellar system of density $\rho(r)$ is given by:
\begin{equation}
f(\varepsilon) = { {1} \over { \sqrt{8} \pi^2 } } 
{ {d} \over { d \varepsilon } }
\int_0^\varepsilon { { d \rho } \over {d \Psi} } { { d \Psi } \over 
{ \sqrt{\varepsilon - \Psi} } },
\end{equation}
where  $\varepsilon$ and  $\Psi$  are respectively  the  relative energy and
relative potential, and that the  distribution function derived in this  way
is physical if and only if the expression
\begin{equation}
\int_0^\varepsilon { { d \rho } \over {d \Psi} } { { d \Psi } \over 
{ \sqrt{\varepsilon - \Psi} } },
\end{equation}
is an increasing function of $\varepsilon$.  Given that the potential of the
homogeneous sphere is $\Phi(r)=-2 \pi G \rho (a^2 - {1 \over 3} r^2)$, it is
clear that  this model   does  not  satisfy the  above  condition,  so   the
homogeneous   sphere  cannot be  a     steady-state  configuration  for    a
collisionless self-gravitating system, as is intuitively obvious.

We   were unable   to  solve  equation~5   analytically to   find the  least
concentrated collisionless  self-gravitating spherical spherical system that
fulfills the    condition of  equation~6; however, by    trial  and error we
discovered that the following, almost homogeneous, density profile does give
a physical distribution function (the  profile is simply the central regions
of a Gaussian):
\begin{equation}
\rho(r) = \rho_0 { { e^{ - ( r r_s/r_l )^2 } - e^{ -{r_s}^2 } } \over 
{ 1 - e^{ -{r_s}^2 } } }.
\end{equation}
Here $\rho_0$ is the central density, $r_s$ is a  scale radius, and $r_l$ is
the  limiting radius  of  the   system.  In   Figure~7, this profile    (for
$r_s/r_l=1$) is compared to our very low concentration King model K17.

Ideally, we would have modeled  both the dark  matter and stellar components
as collisionless dynamical systems and  simulated both with the N-body code.
One  would then have  been able to compare  the modeled stellar component to
the observed kinematics and observed light profile.  However, this is a very
difficult numerical task, as,  in order to  avoid mass segregation, all  the
particles in the simulations must  have the same  mass.   Since the mass  to
light ratio  of \sgg\ deduced  by  IWGIS is $M/L \sim   100$, the number  of
particles  in the combined stellar and  dark matter simulation would have to
be  $\sim 10^6$  in  order to  obtain  a  useful  resolution in the  stellar
component. So in the present contribution, we only consider the dark matter,
which must be the principal contributor to the potential,  if it is to solve
the problem of the survival of this dwarf galaxy.

We experimented releasing models on the `a' orbit with a range of structural
parameters into   the fixed Milky Way  potential.   It  is  uninteresting to
describe these  pure dark   matter models  at  any  length, since  there  is
relatively little  to learn from their end-point  structure given that these
cannot be compared to observations.  We were contented to find that one such
model,  of $N=8000$  particles, and  with parameters  $M =   1.2 \times 10^9
\msun$, $r_s=1\kpc$,  $r_l=1\kpc$, managed to  survive largely intact at the
end  of the simulation ($42$\%  of the particles  remained inside $T_{ok}$).
Contrary  to what  was found  in  the simulation   of  the King  models, the
present-day minor   axis  half-mass  radius  for  this  dark  matter  model,
$R_{HB}=0.7\kpc$, is more extended than the stellar component.

Thus one  can construct dark matter  models that survive the Galactic tides.
The last example gives a flat, extended dark matter  profile, as required by
the analytical dark matter halo model proposed by  IWGIS. However, given the
absence of a luminous component in  the simulations, at  the present time we
are unable to make a useful comparison to observations.

\section{Discussion}

\subsection{The assumptions}

\subsubsection{The choice of Galactic potential}

At present, the constraints  on the Galactic mass distribution, particularly
in the  vertical direction, are  not very  tight, so the  range of plausible
Galactic mass models remains  large (\markcite{deh97}Dehnen \& Binney 1997).
This means that the particular choice for the Galactic potential model could
have biased the survivability of the models. To check this possibility, one
of the robust dwarf galaxy models (K17-a) was  simulated under the influence
of three models for  the Galactic  potential.   The mass fraction within  $5
\kpc$ of  the center of  the model at the  end of  the simulation is $77$\%,
$86$\% and        $56$\%    (\cf  Table~1),    in,      respectively,    the
\markcite{joh95}Johnston  \etal\ (1995) potential, the \markcite{eva94}Evans
\& Jijina  (1994) potential with $q =  1$, and the  \markcite{eva94}Evans \&
Jijina (1994) potential with $q = 0.9$.  Thus the disruption rates, in these
quite  different, Galactic potentials are   similar, which suggests that our
results    on the survivability   of  \sgg\ are  not  very  sensitive to the
differences in {\it plausible} Galactic potential models.
\footnote{The orbit  in   the flattened  halo  has a  smaller periGalacticon
passage, which leads to more rapid disruption.}

\subsubsection{Evolutionary changes in the Galactic potential}

It has also been  assumed that the  Galactic potential is static. Clearly in
the early stages of  the formation of the  Milky Way, the Galactic potential
must have changed quite considerably.  However, the thinness of the Galactic
disk  argues  strongly (\markcite{tot92}T\'oth  \& Ostriker   1992) that the
Galaxy  did not  suffer  any  major  mergers   since the  formation of  that
component, more  than     $9.5 \Gyr$  ago   (\markcite{osw96}Oswalt   \etal\
1996). Another way  that the Galaxy may  have grown is  by  the accretion of
many  small primordial clumps such as  dwarf  galaxies; as discussed further
below, at  least   from spheroid  stars  in  the  Solar  neighborhood   halo
(\markcite{una96}Unavane  \etal\ 1996),  one may  rule  out  any substantial
amount  of accretion   in this  form  onto the  Milky  Way.   So  though the
assumption that the  Galactic potential has  not  changed substantially over
the last $12 \Gyr$ is optimistic; it is probable that,  for a large fraction
of  that time,  it has  remained approximately  constant.  If  the potential
gradients were substantially lower in the past, the  tidal stresses on \sgg\
would have been lower, and the simulations will have been biased to finding
shorter survival times.

\subsubsection{The effect of the Magellanic Clouds}

We have  assumed that the Milky Way  provides the only  significant external
gravitational force  field on \sgg.  However, it  is  conceivable that other
members of the Galactic satellite system may play a r\^ole in its evolution.
In  particular, the  Large and Small   Magellanic Clouds are massive  enough
(respectively, $6  \times 10^9 \msun$,  \markcite{mea88}Meatheringham \etal\
1988; and $\sim 1.2 \times  10^9 \msun$, \markcite{lin95}Lin \etal\ 1995) to
affect   the  orbit and the  structure   of \sgg\  if  they   happen to pass
sufficiently close to it.  Even though the orbital plane  of \sgr\ is almost
perpendicular to  that of the Magellanic  Clouds, such a close encounter may
indeed occur,  given  that the  apoGalactic distance  of  $\simgt 60   \kpc$
deduced  above for the orbit of  \sgg\  is approximately coincident with the
present Galactocentric distance of the Magellanic clouds.

To  investigate  this possibility,   we  performed a simulation  with  three
particles, representing \sgr,  the LMC and  the SMC.  The kinematics  of the
Magellanic Clouds were taken from \markcite{kro97a}Kroupa \& Bastian (1997),
as  determined   from a  mean   of Hipparchos    proper motions and  earlier
measurements. \Sgg\ was assumed to move on the orbit `a', and to have a mass
$M = 10^9 \msun$.  The N-body code was used to integrate  the motions of the
three dwarf galaxies  backwards in time,  under the influence of the  static
potential of  Johnston \etal\ (1995), with  dynamical friction reversed. The
forces were computed by direct summation, and a  softening length of $1\kpc$
was adopted.  With the above-stated parameters, the orbit of \sgg\ is indeed
altered by the Magellanic  Clouds, but the effect  is small, as can be  seen
from  Figure~8. In Figure~9,  the  distance between  \sgg\ and  each  of the
Magellanic Clouds  is  plotted as a function  of  time. Evidently, with  the
above parameters, the impact parameter never becomes smaller than $20\kpc$.

To quantify  the deviation on   the orbit of  \sgr\   due to the  Magellanic
Clouds, we  performed 1000 simulations, each  with  a different, but equally
likely, realization of the kinematic  starting parameters of the  Magellanic
Clouds given  the published errors (which  we  assume to  be Gaussian).  For
each of these  simulations, we  computed the  final  (i.e.  the  primordial)
radial      period  (defined as  the   time     taken from  apogalacticon to
perigalacticon  and back).  The resulting  distribution of radial periods in
the   ensemble of simulations  is displayed  in Figure~10; the  mean of this
distribution, at  $1.09 \Gyr$, is very   close to the unperturbed  period of
$1.08\Gyr$, while  the rms spread  is $0.031\Gyr$.  The  minimum and maximum
periods  found were  $0.977\Gyr$ and  $1.37\Gyr$.  In Figure~11  we show the
distribution of minimum  impact distances between  \sgg\ and  the Magellanic
Clouds.  The probability of encounters with impact parameter less than, say,
$5\kpc$,  is $4.3$\%.  Thus, it is  very unlikely that the Magellanic Clouds
alter significantly the orbital path of \sgg, though there is a small chance
that they may affect its internal dynamical evolution and its star-formation
history.  (\Sgg\ may similarly affect the  Clouds).  The effect of the extra
gravitational perturbation will undoubtedly be to increase tidal disruption,
and so  diminish the  lifetime  of \sgg.  The  fact   that we neglected  the
Magellanic Clouds in the  disruption simulations therefore poses  no problem
to our aim of placing a lower bound on the survival time of \sgg.

\subsubsection{Accretion onto \sgg}

It was assumed  that there has been  no significant accretion onto \sgg. Yet
\sgr\ contains a mix  of  stellar populations,  ranging from  relatively old
stars -- many   RR~Lyrae stars are observed (\markcite{mat95b}Mateo   \etal\
1995, \markcite{ala96}Alard 1996) and at  least one of its globular clusters
is as old as the oldest Galactic halo clusters (\markcite{ric96}Richer \etal
1996, \markcite{cha96}Chaboyer,   Demarque  and   Sarajedini  1996)   --  to
intermediate  age  stars  --  several  Carbon   stars have  been  identified
(\markcite{me94}Ibata  \etal 1994,  \markcite{me95}1995).  It  is unclear at
present   whether this complex   star-formation  history  is the  result  of
internal enrichment   or  accretion from beyond    the dwarf.  However,  the
accretion of a significant  amount of matter   onto the dwarf  requires very
special circumstances
\footnote{For instance, one can imagine  an accretion scenario where  clumps
of gaseous material,  driven out from the  outer Galactic disk by SNe winds,
attain  velocities similar to the orbital  velocity of \sgg. As \sgg\ passes
close to such clumps, depending on relative velocities, some material within
a Bondi--Hoyle radius could be accreted.},
so   it is more conservative to   assume that  only  internal processing has
produced the observed metal-rich stellar populations, and that the dwarf has
only lost material.

\subsubsection{Effect of a collisional component}

We  have  also  assumed that  the dwarf    was always  entirely  composed of
collisionless particles. Clearly,  a substantial gaseous component must have
been present in  the dwarf at each stage  of star-formation. However, we can
expect that most such material would  be stripped from  the dwarf during its
collisional encounters  with the Galactic disk.  If  there was no significant
accretion, the periodic removal  of gas from \sgg\  would have the effect of
lowering the   total mass, thereby   reducing  its  survival   time. By  not
including this effect, we have under-estimated the disruption rate, which is
not a worry in terms of our aim of determining a lower bound to the survival
time.

\subsection{Survival of primordial galaxy fragments}

One  of   the  most   significant  consequences of   the   failure   of  the
mass-traces-light models to reproduce the  observations is that the  merging
fragments  that made the  Galaxy probably had  a radially increasing mass to
light ratio, so that the stars were more centrally concentrated and the dark
matter more extended. Detailed numerical simulations are required to explore
this question further, but one may expect such a structure to initially lose
almost exclusively  dark matter,  with stars being  lost  only  in the  last
stages of disruption. If dynamical friction causes significant orbital decay
of the merging clumps, the luminous matter would be deposited more centrally
than the dark matter in the global potential well,  naturally giving rise to
a radially increasing mass to light ratio in the halos of large galaxies.

We have shown that,  even in the  absence  of a dark matter  component, some
dwarf  galaxy models  (those a  small initial  half-mass radius) may survive
almost intact in the Galactic halo, even on such short period orbits as that
deduced for \sgg.  Furthermore,  the range  of allowable initial  structural
parameters of the  models is quite  large.  One may  deduce from these facts
that a  significant fraction of  the primordial population of dwarf galaxies
with initial half mass radius $\sim 1/3$ of  that of \sgg\ managed to survive
until the present day; this holds as long as the orbits of the population of
proto-dwarf galaxies were  not  significantly  more biased towards   radial
orbits, where smaller perigee distances give rise to faster disruption.  The
observed paucity of such  galaxies in the  Milky Way halo therefore provides
strong  evidence that the   primordial population was   not  numerous.  This
conclusion is consistent with the findings of \markcite{una96}Unavane \etal\
(1996), who, from counts  of Solar neighborhood  spheroid stars blue-ward of
${\rm B-V = 0.4}$, put a limit on the merging rate into the Galactic halo of
40 Carina-like dSph, or 5 Fornax-sized dSph in the last $10 \Gyr$.

However,  some cautionary   remarks  on the  \markcite{una96}Unavane  \etal\
(1996) technique  should  be  made.   In all our   simulations, the material
disrupted from  the models is  confined to long streams  that persist for at
least $12 \Gyr$.  The mass density along the streams is inversely related to
the  orbital  velocity,  so most  of    the disrupted  mass  lies  at  large
Galactocentric  radii.    Also, a   negligible  fraction  of  the  disrupted
particles attain orbits with  significantly lower perigee distance than  the
parent satellite. For these reasons, one may expect disrupted dwarf galaxies
to   contribute  a   negligible  fraction  of    their mass   to  the  Solar
neighborhood. Only stars disrupted from  those dwarf galaxies that had small
periGalactic distances,  but long enough   orbital periods not to have  been
destroyed  at the earliest epochs,  will be seen  near the Sun; furthermore,
the kinematics of such stars will likely be quite different from that of the
spheroid.

Recently,  \markcite{kro97b}Kroupa (1997) has argued that  the  high mass to
light ratios inferred for dSph  galaxies are incorrect, because the measured
high velocity dispersion arises from viewing tidal debris along a disruption
orbit  that is  approximately parallel to   the line of  sight.   If so, the
present  day  dSph are only   small  remnants containing  $\sim  1\%$ of the
original mass of these galaxies.  Though this may a feasible explanation for
the velocity dispersion of some  of the small  dSphs, this suggestion cannot
possibly hold for  \sgg, as the  constraints on disrupted  material from are
quite strong.  If \sgg\ at the present day contained only  $\sim 1\%$ of the
original galaxy, we would expect to see $\sim  400$ globular clusters on the
great circle of its orbit. This is clearly inconsistent with observations.

\subsection{Predictions of the models}

All of the simulations  give rise to  streams of tidally  disrupted material
that follow the  orbital path of  the remnant quite closely.  The  disrupted
fraction   always exceeded  $15$\%,  though  we stress  that  this is simply
indicative, since we have  not managed to make any  model fit the  available
data. So we may expect to find a  sizeable fraction of the stellar component
of \sgr, including perhaps one or more globular clusters, stretching along a
ring around the sky.   An interesting  feature of  the streams is  that they
become narrow near perigee and broad at apogee.  It will be very fruitful to
detect such material, as its kinematics  could provide a very sensitive test
of the Galactic potential   gradients; we will   discuss this further  in  a
subsequent contribution.

As  discussed in IWGIS,  the mass range  of  Sagittarius dwarf galaxy models
that contain dark matter, suggest that the Milky Way itself is significantly
affected by the passage of \sgg; certainly the tidal forces on the Milky Way
due to \sgg\ are substantially larger than those due to the Large Magellanic
cloud,  which has previously   been invoked as a   possible perturber of the
\ion{H}{1}  disk  (\markcite{wei95}Weinberg 1995).  Thus, the star-formation
history of the Galaxy  may have been influenced by  this dwarf.  To quantify
the  picture, it will  be necessary to  determine  the mass of  the dwarf as
accurately as possible.   Such a study is possible  by combining an accurate
photometric profile  (which has  not as yet  been  obtained) with the extant
velocity  dispersion  profile.   This   information  determines   both   the
present-day dark matter profile and the present total mass.  Detailed N-body
simulations, with a stellar component embedded in the dark matter, will then
have to be performed to check the robustness of  the models and whether the
end-point structure  can be made  to  be consistent  with observations. This
gives an opportunity to determine the distribution of dark matter in a dwarf
galaxy.

\section{Conclusions}

Comparison  of  the observed velocity  profile  to the simulations presented
above indicates that \sgg\ has a  short period orbit,  with radial period $T
\sim 0.7 \Gyr$.   It  is found that  any  reasonable model of  the  internal
structure of the dwarf galaxy, where light traces mass and where $M/L \simlt
10$, either does not  survive the tidal interaction with  the Milky  Way, or
has a minor  axis half-mass radius that  is  inconsistent with observations.
Thus, it is not  possible to understand  the present  existence of  \sgr\ if
most of its mass  is in  the form of   stars.  However, this problem may  be
solved if Sagittarius, and by  implication, other dwarf spheroidal galaxies,
have a radially increasing mass to light ratio, as suggested by IWGIS.  This
analysis supports the mass to light  ratio determination of IWGIS, who found
$M/L_{\rm global} \sim 100$.

This conclusion should  still  hold if there   is internal rotation  in  the
dwarf.  Nor should the conclusion be affected significantly by the choice of
Galactic potential model.  The  perturbative effect of the Magellanic Clouds
was  considered; given  current estimates of  their  kinematics, there is  a
negligible chance   that they could have  altered  sufficiently the orbit of
\sgg\ to account for the short period deduced above.

Further numerical  work is  required to  construct  a  fully self-consistent
model in which  both the stellar and dark  matter components are present and
can reproduce the observations.

\bigskip

We are very  grateful to Derek  Richardson for kindly letting  us use of his
{\tt box\_tree} code and related  plotting routines, and especially for many
illuminating email conversations. We also   cordially thank G.  Fahlman,  K.
Menon, H.  Richer, D.    Scott and G.   Walker  for  generous loan  of their
computer  resources.  RAI  expresses    gratitude to the  Killam  Foundation
(Canada) and to the Fullam Award for support.  \vfill\eject

\clearpage

\begin{figure}
\psfig{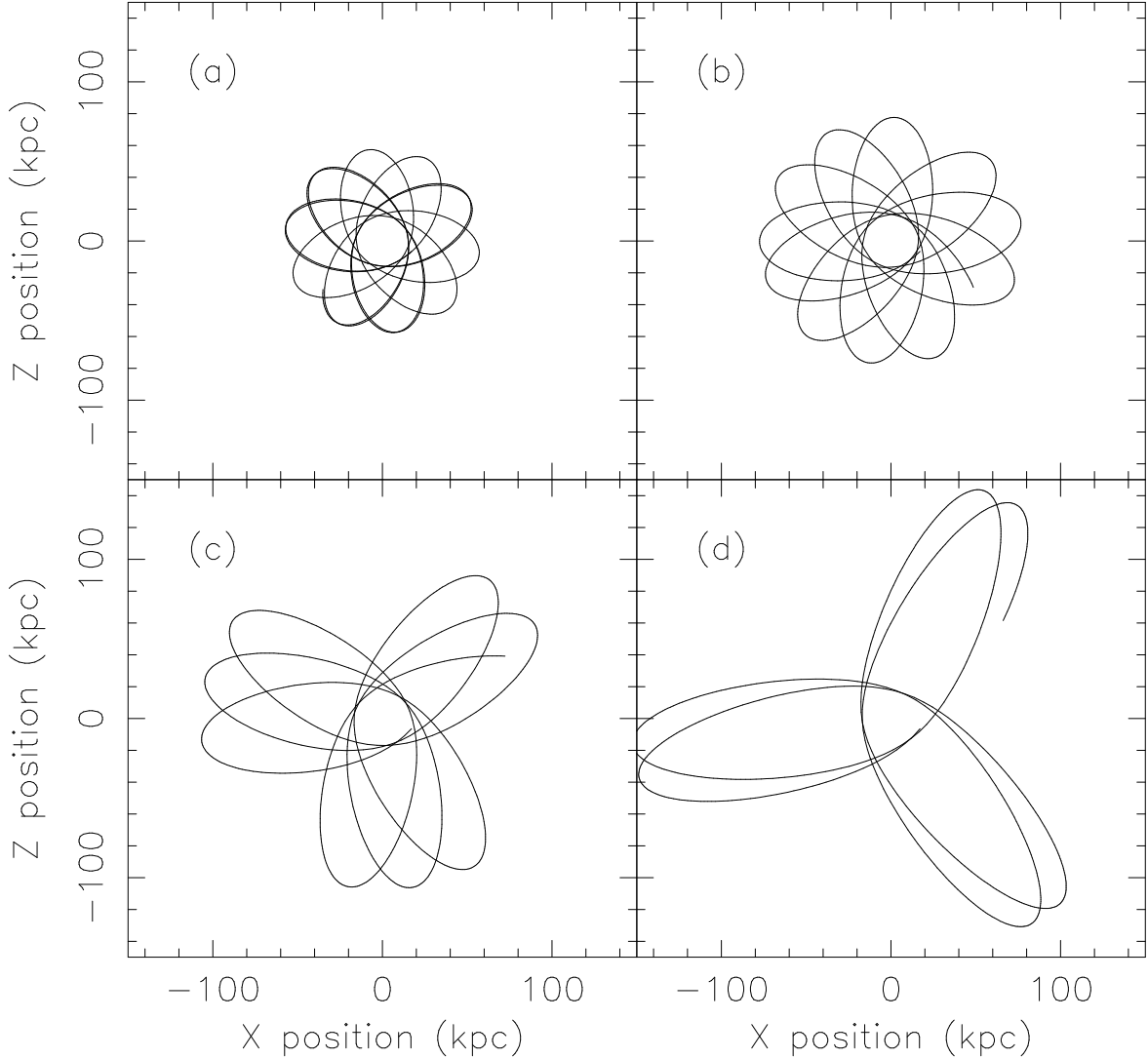}
\figcaption[Ibata.fig01.ps]{ 
These diagrams show  the $x$--$z$ structure of  the orbit of a test particle
(\ie negligible dynamical friction)   integrated backwards in time for  $12
\Gyr$  under the influence  of   the  Galactic  potential  detailed in   the
text. The orbit in panel `a' is derived  from the best  fit to the kinematic
and distance observations. This orbit has a total velocity of $|v_{com}|=332
\kms$ at the present position of \sgg.  However, for reasons detailed in the
text, we also explore the behavior of the dwarf galaxy structural models on
the longer period  orbits  shown in panels `b',   `c'  and `d',  which  have
$|v_{com}|= 362 \kms, 392 \kms$ and $422 \kms$, respectively.}
\end{figure}
\clearpage

\begin{figure}
\psfig{figure=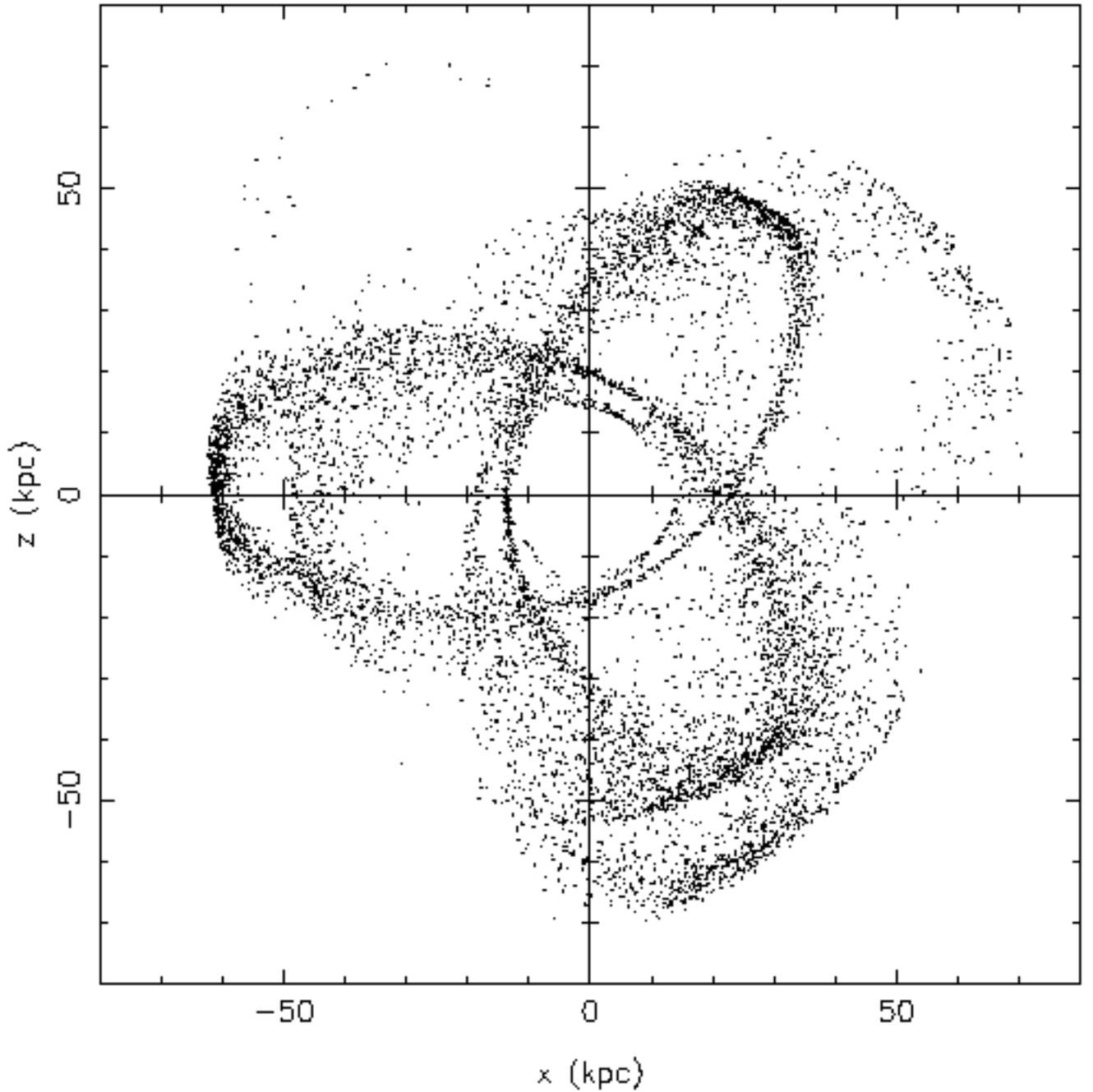}
\figcaption[Ibata.fig02.ps]{The $x$--$z$ plane structure  of the remnant  of
the  dwarf   galaxy  model K1-a,  whose  initial   structure  fits  best the
present-day  observations of   \sgg, is  shown    at the  end-point of   the
integration.  All traces of a central concentration have vanished.}
\end{figure}
\clearpage

\begin{figure}
\psfig{figure=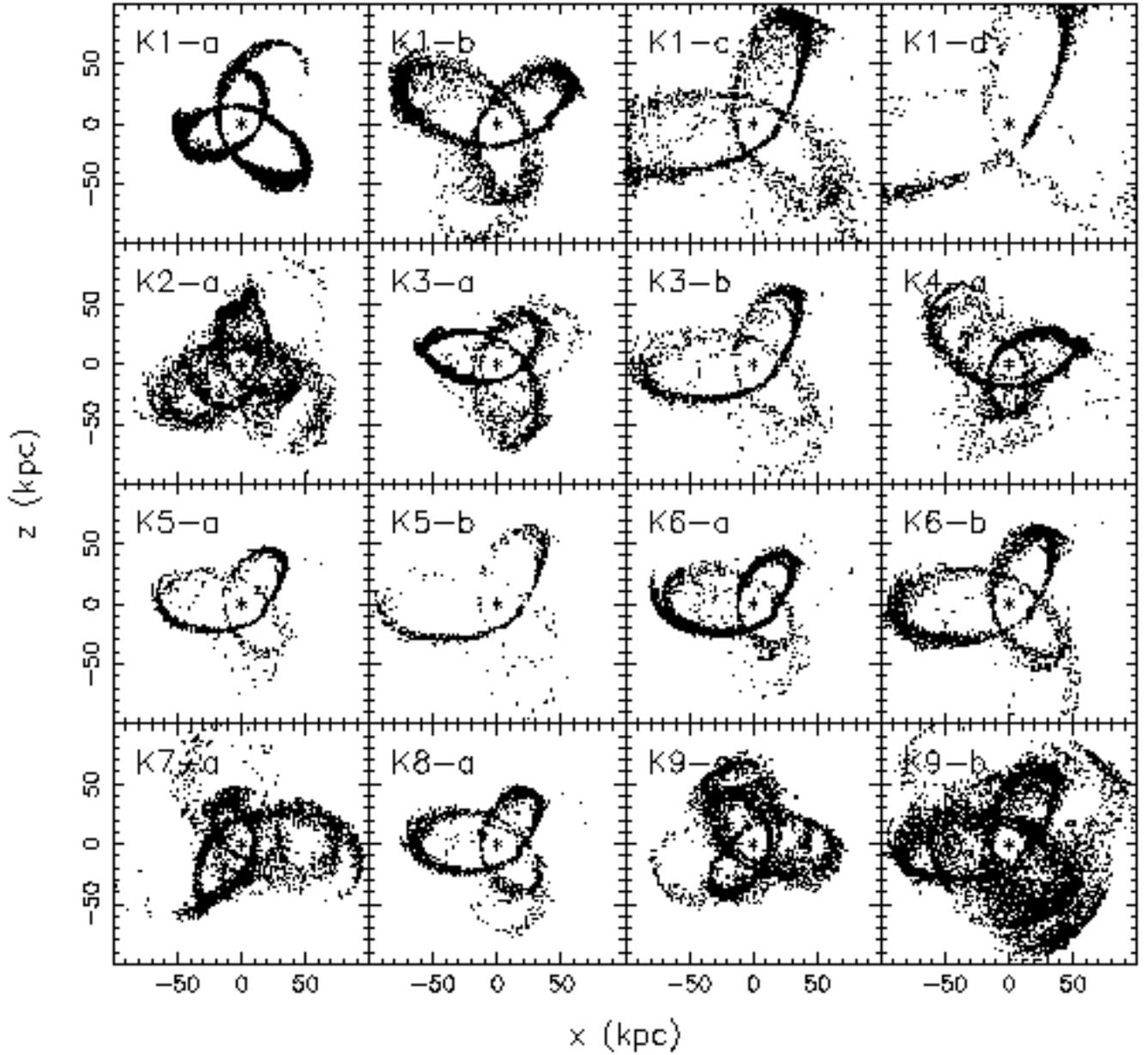}
\figcaption[Ibata.fig03.ps]{The structure of all  the King model simulations
(except the two models  simulated with  the  Galactic potential of  Evans \&
Jijina 1994)  is displayed at  the  simulation time $T_{ok}$.  Each panel is
marked with   the  model identification  label,  as given  in  column (1) of
Table~1.}
\end{figure}
\clearpage

\begin{figure}
\psfig{figure=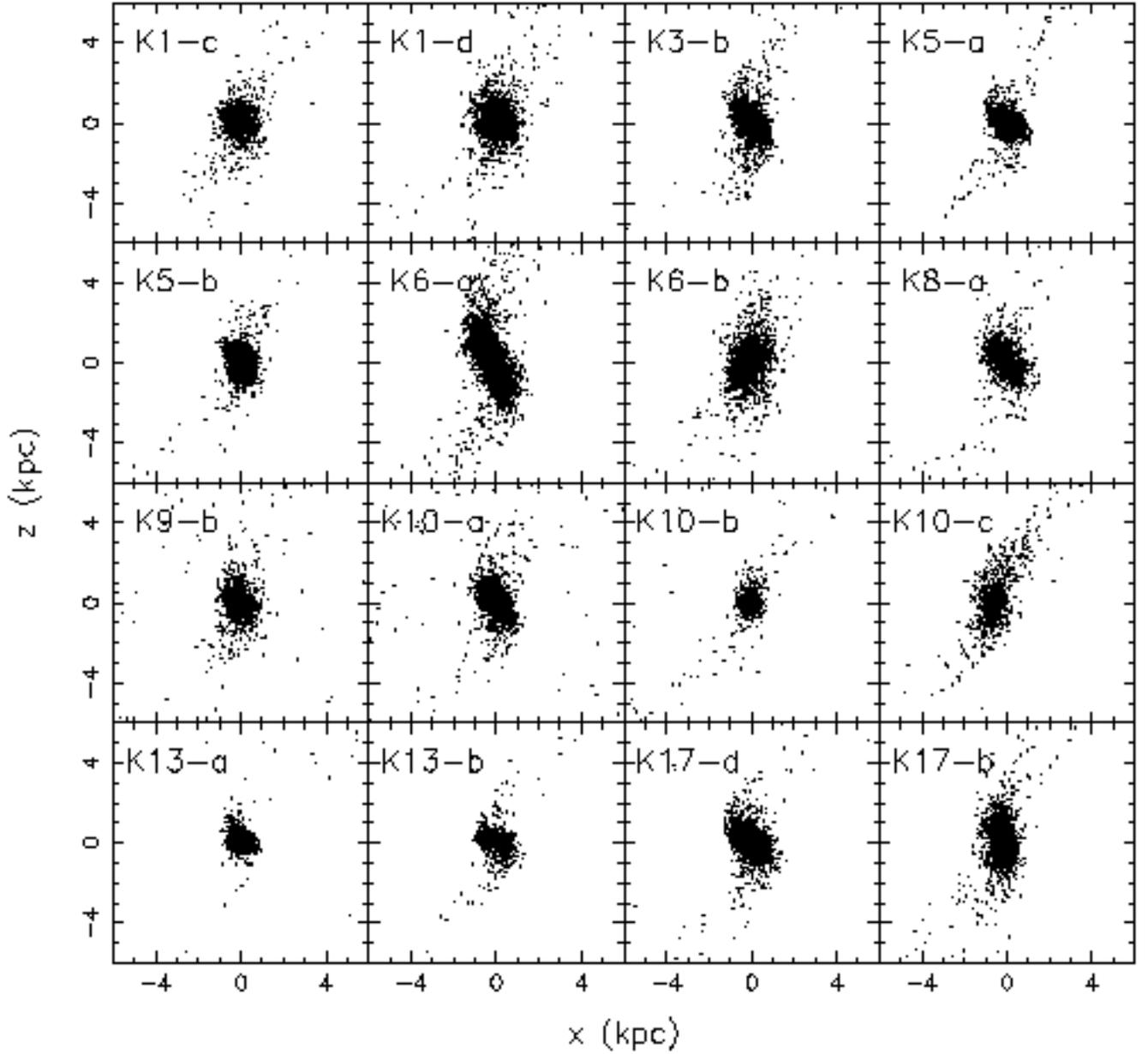}
\figcaption[Ibata.fig04.ps]{A close-up picture of the end-point structure of
all the models  in Figure~3 that managed to  retain a central  concentration
until the end of the simulation.}
\end{figure}
\clearpage

\begin{figure}
\psfig{figure=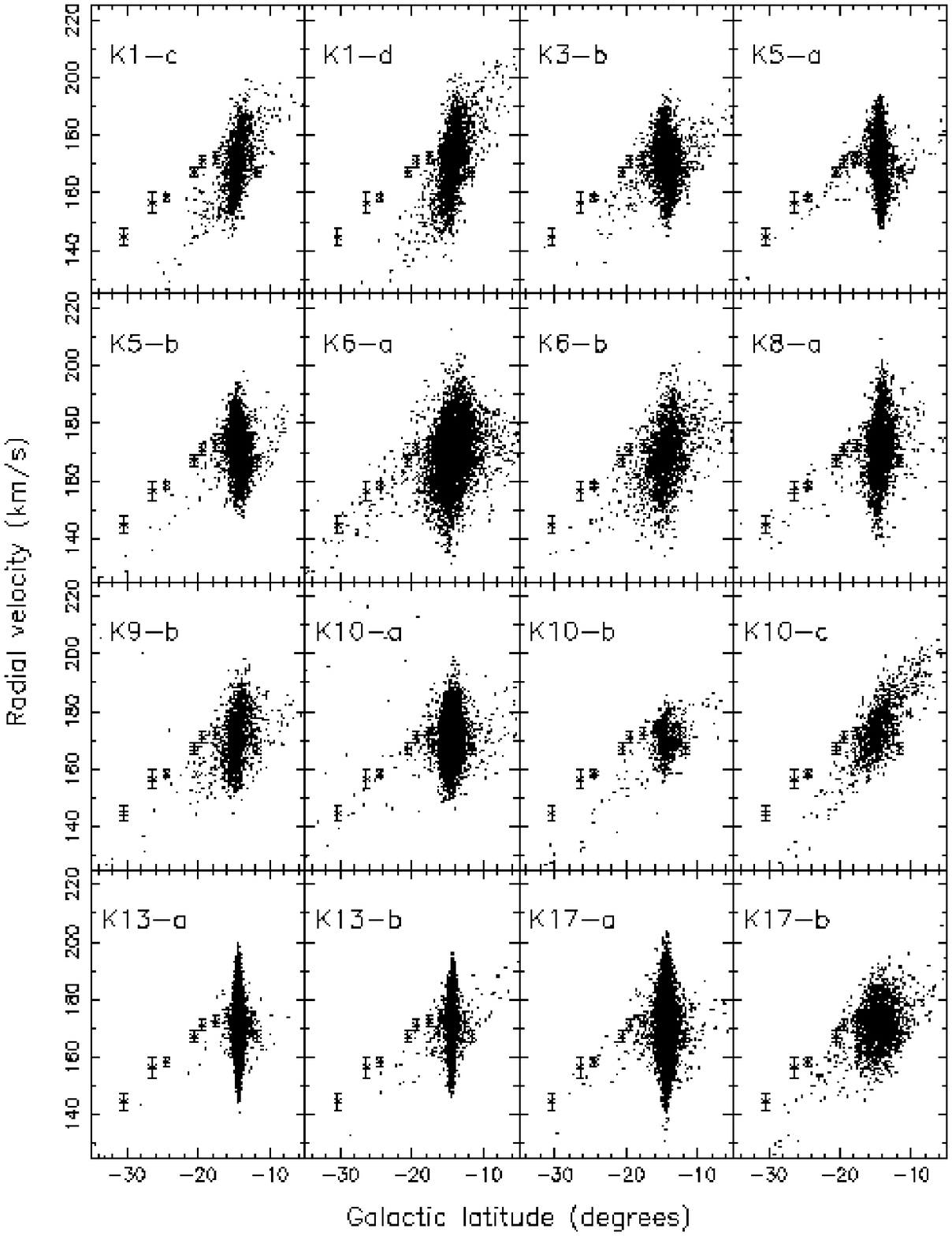}
\figcaption[Ibata.fig05.ps]{The  observed     radial velocity  gradient   is
compared to   the  models  displayed in    Figure~4.   For  the purpose   of
constructing this diagram, the simulations were halted as  the center of the
remnant passed   $b=-14.5^\circ$,  (which we assume to    be  the center  of
\sgg). Comparing the velocity profile of the material that has been torn off
the model, one can clearly see that the long  period orbits `b', `c' and `d'
give a poor match to the observations. The velocity profile of the models on
the `a' orbit fit the observations much better.}
\end{figure}
\clearpage

\begin{figure}
\psfig{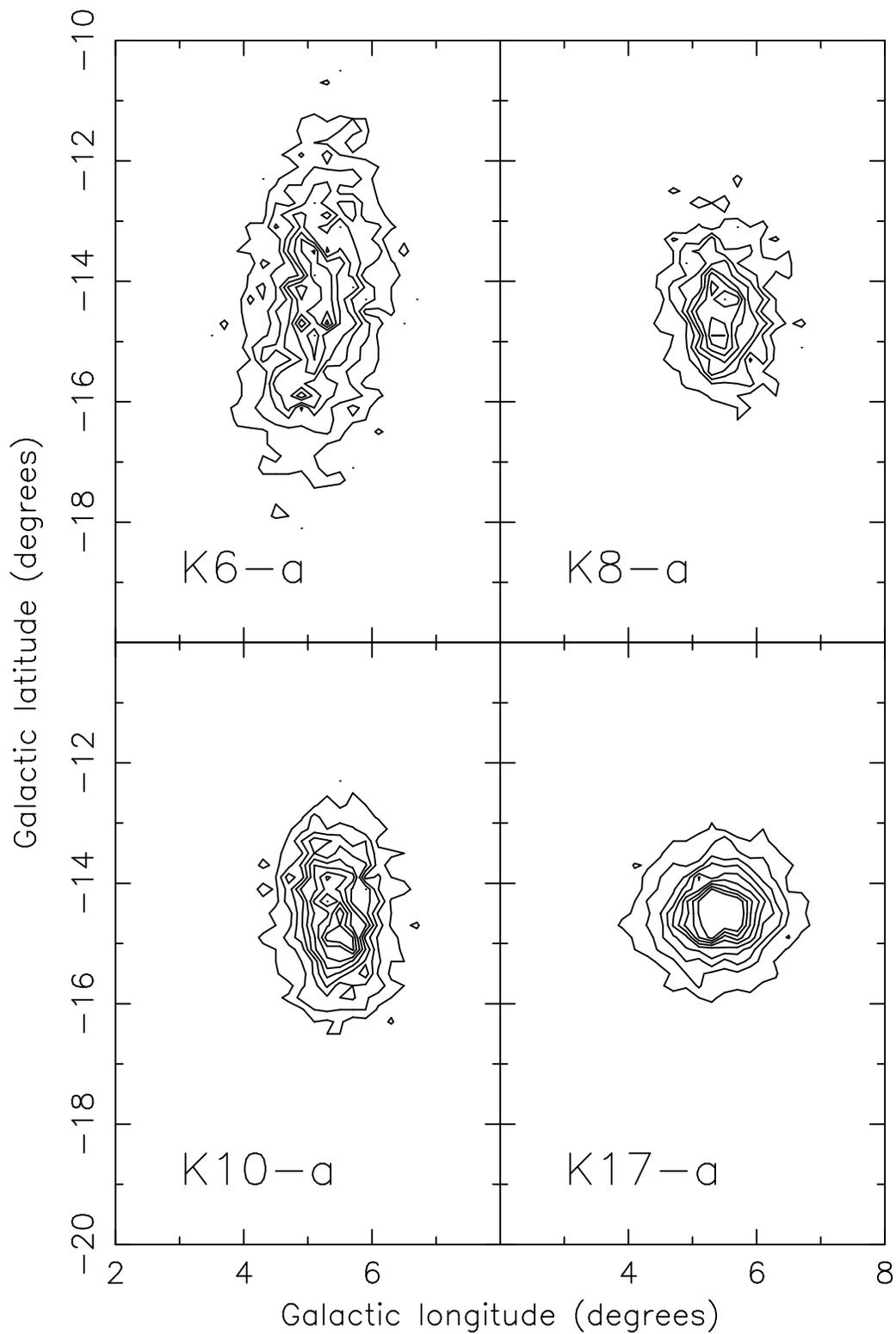}
\figcaption[Ibata.fig06.ps]{The  final  surface number  density of particles
from  four  robust  King models  that  give  a  good representation   of the
kinematic observations, are displayed. However, all four models have a minor
axis half-mass  radius $R_{HB} < 0.21  \kpc$, inconsistent with the observed
value of $R_{HB} \sim 0.55 \kpc$.}
\end{figure}
\clearpage

\begin{figure}
\psfig{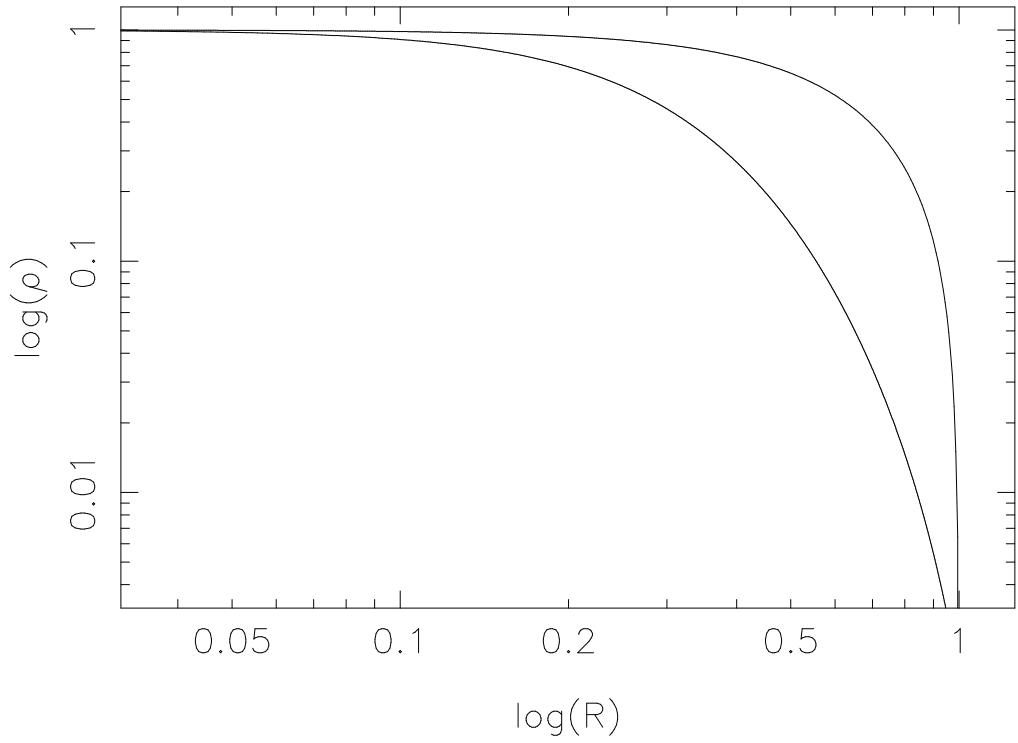}
\figcaption[Ibata.fig07.ps]{The  density profile    of the  Gaussian   model
described in the text (top curve) is compared to  the very low concentration
King model K17 (bottom curve).}
\end{figure}
\clearpage

\begin{figure}
\psfig{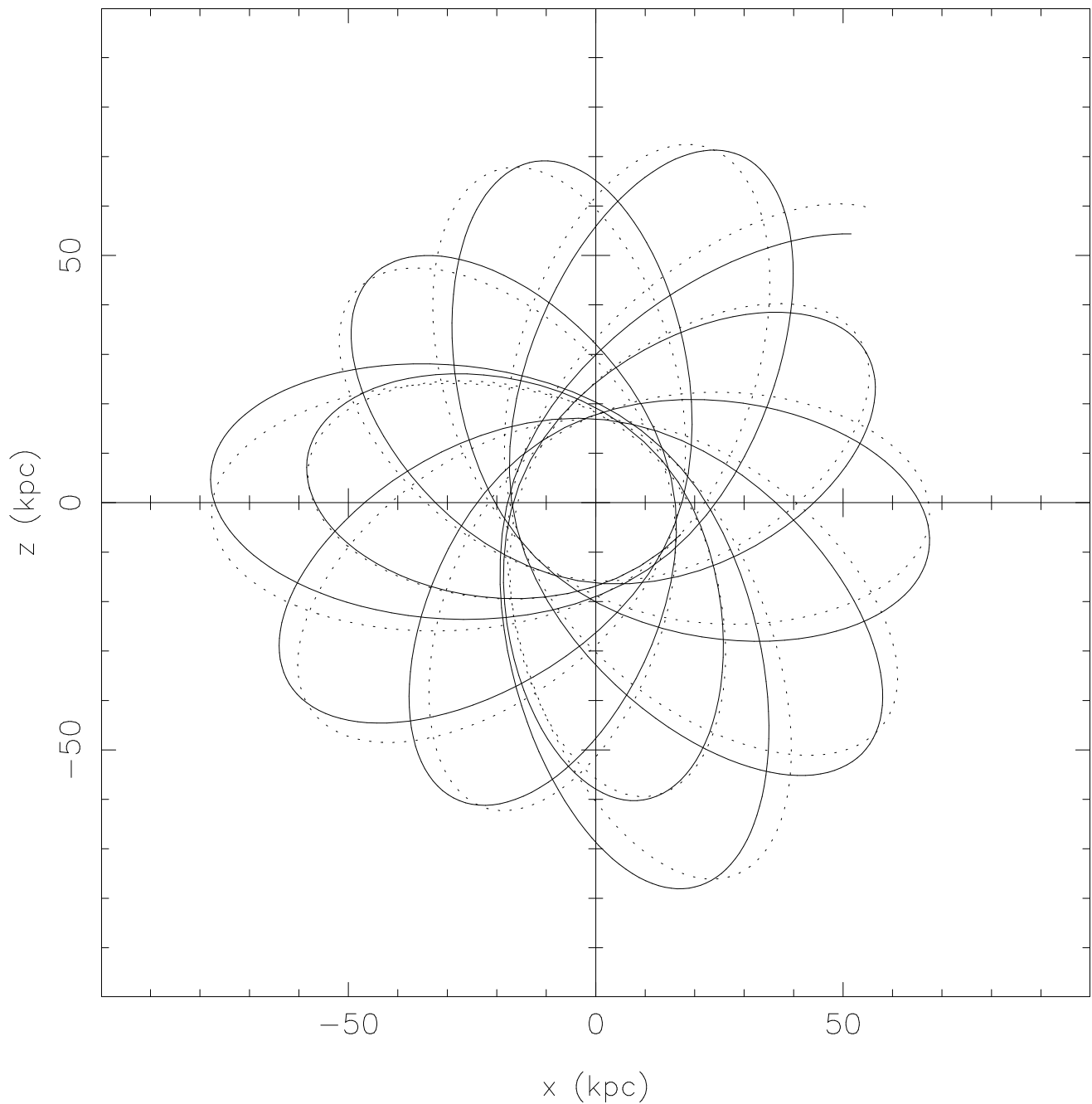}
\figcaption[Ibata.fig08.ps]{The solid line shows  the path, in  the $x$--$z$
plane  of the Galaxy,  of a $M  = 10^9 \msun$  point-like particle that moves
purely under the influence of the assumed  Galactic potential for $12 \Gyr$.
The initial velocity parameters of the particle correspond to those of orbit
`a'.  The dotted line  shows the perturbed  path taken  by the same particle
when  point masses, representing the  Small and Large Magellanic Clouds, are
included into the integration.}
\end{figure}
\clearpage

\begin{figure}
\psfig{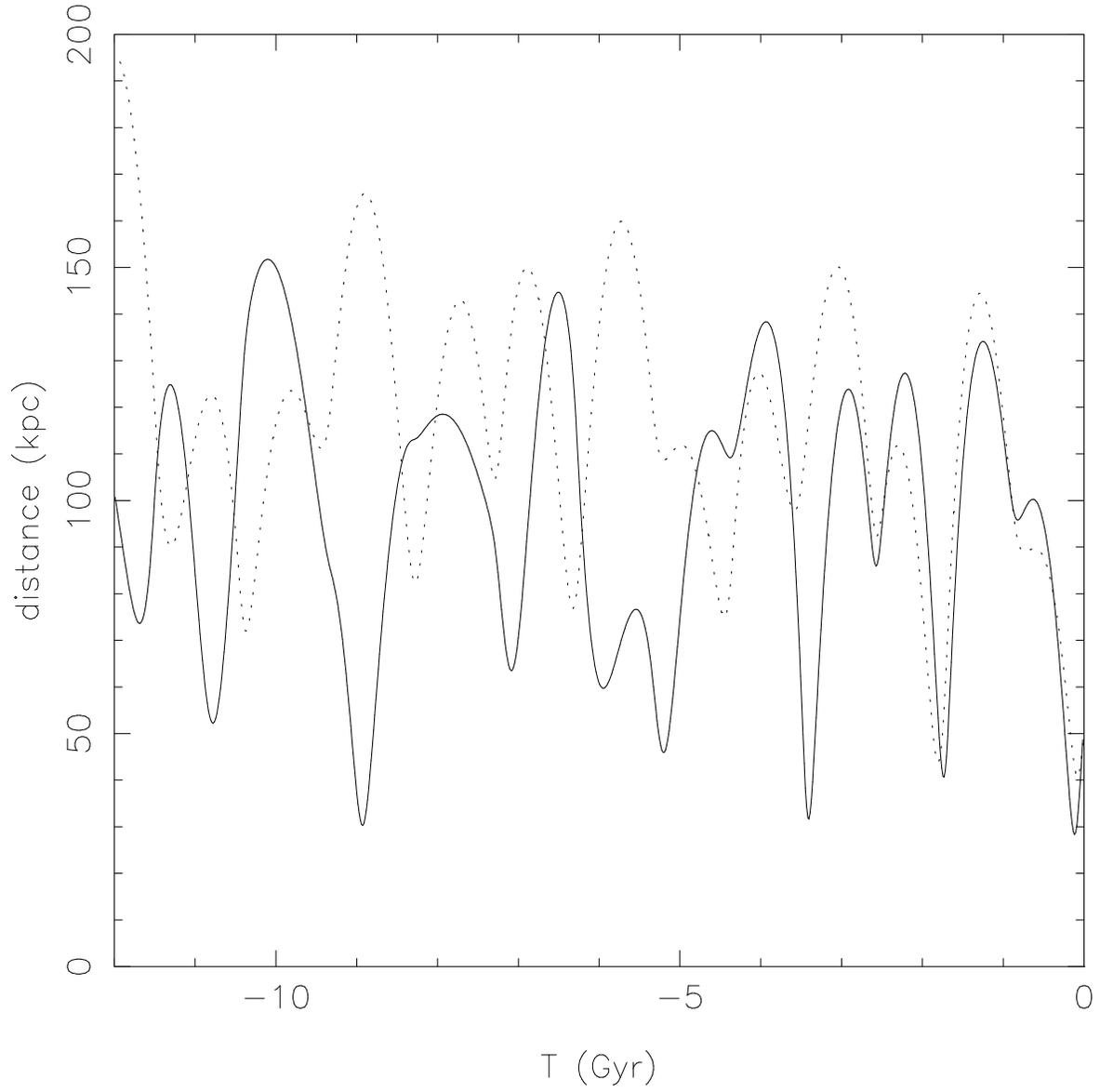}
\figcaption[Ibata.fig09.ps]{The distance between \sgg\ and, respectively, the
LMC (solid line) and the SMC (dotted line) in the simulation of Figure~8 are
displayed as a function of time. The time $T = 0$ corresponds to the present
day.}
\end{figure}
\clearpage

\begin{figure}
\psfig{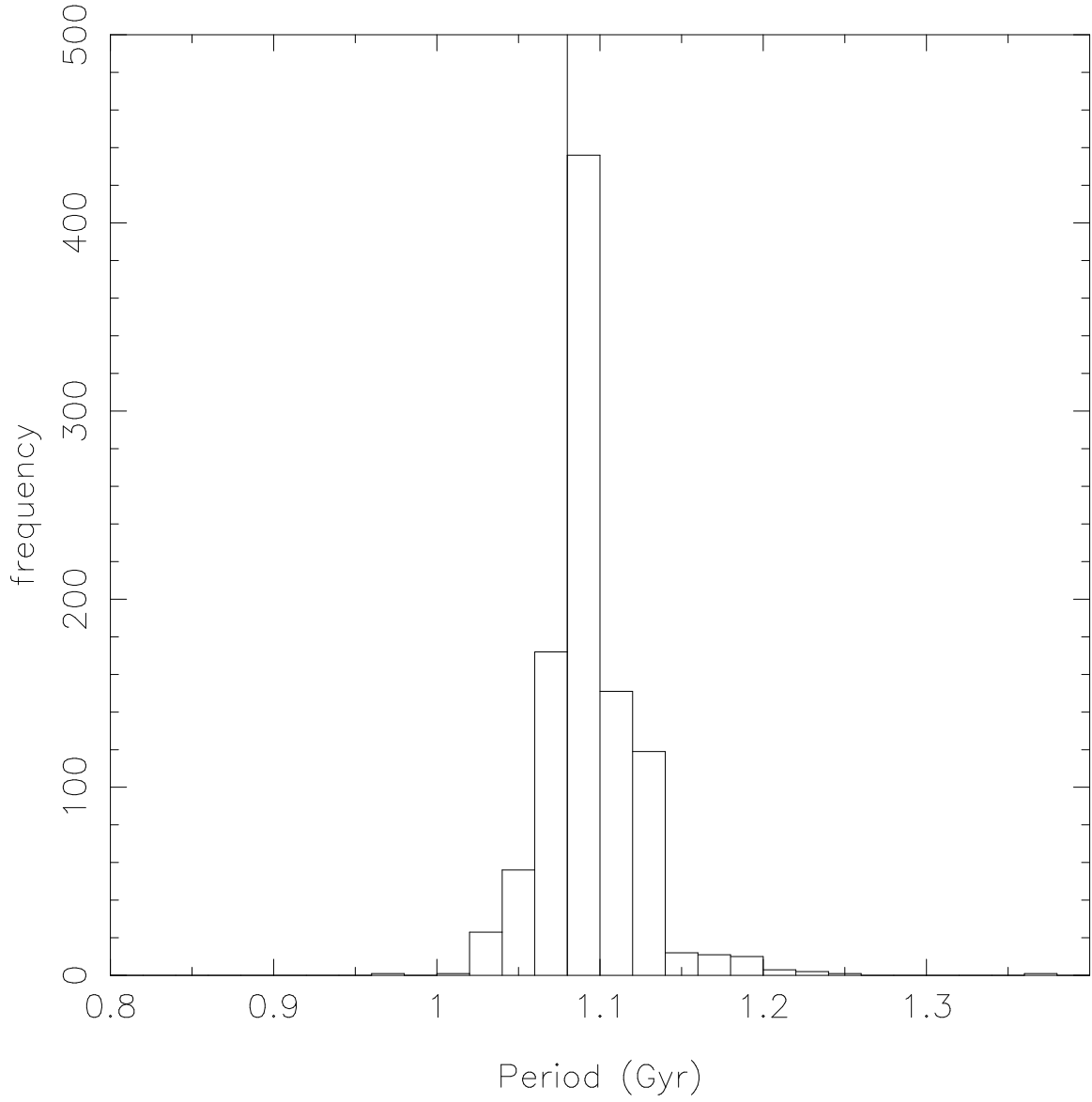}
\figcaption[Ibata.fig10.ps]{The     probability  distribution of     initial
(i.e. primordial) radial periods of \sgg\  in the presence of the Magellanic
Clouds.  Clearly, the period of \sgg\ is  not significantly altered by these
perturbers.}
\end{figure}
\clearpage

\begin{figure}
\psfig{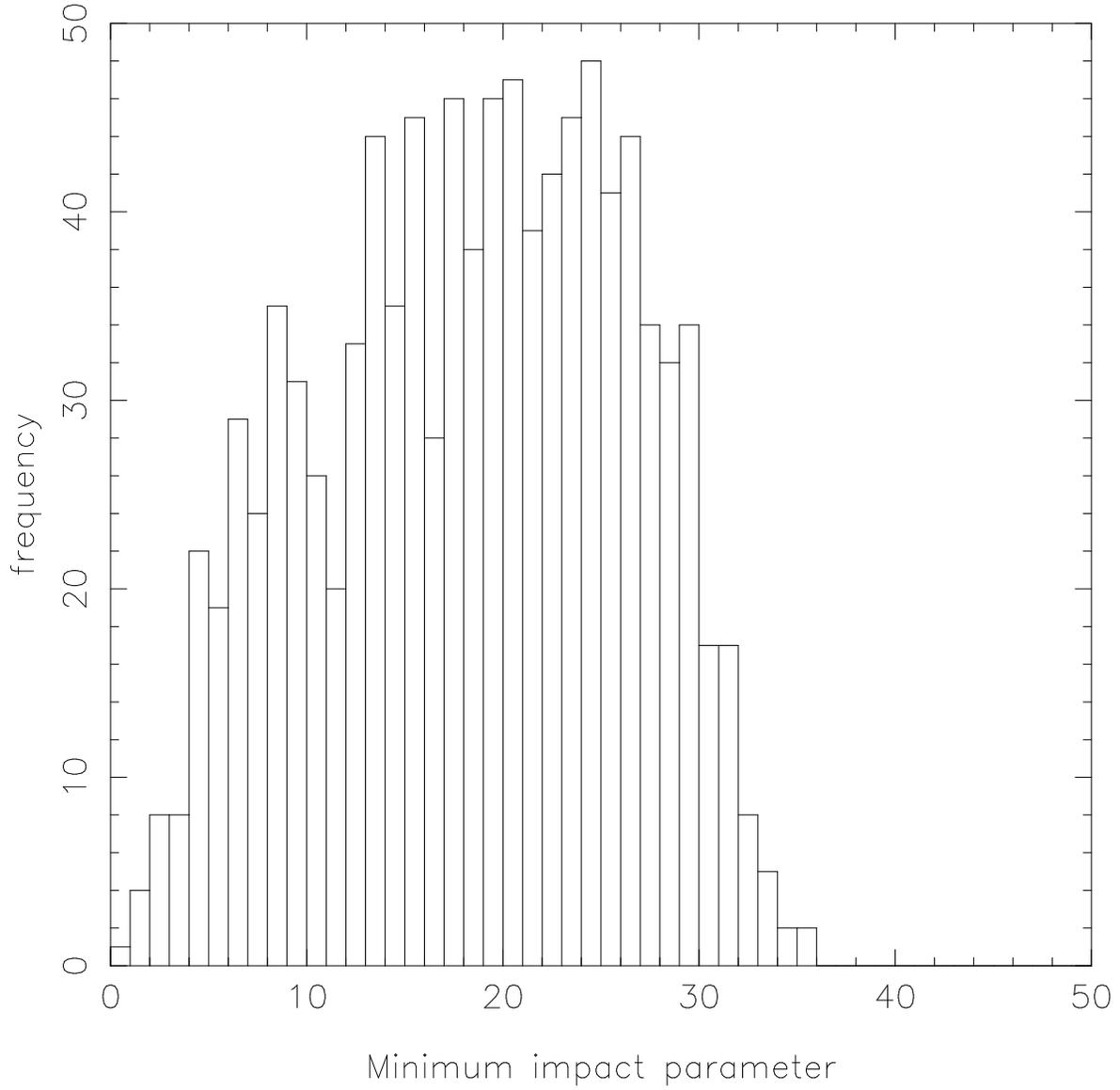}
\figcaption[Ibata.fig11.ps]{The   distribution of minimum impact  parameters
between  \sgg\ and the Magellanic  Clouds.  The chance  of a close encounter
having taken place in the past appears to be significant, but not large.}
\end{figure}
\clearpage


\begin{references}

\reference{ala96}
	Alard, C. 1996, \apj\ 458, L17

\reference{ben90}
	Bender, R., Nieto, J. 1990, A\&A 239, 97

\reference{bin78}
	Binney, J. 1978, \mnras\ 183, 501

\reference{bin87}
	Binney, J. \& Tremaine, S., 1987 `Galactic Dynamics' (Princeton 
	University Press, Princeton)

\reference{cha96}
	Chaboyer, B., Demarque, P. \& Sarajedini, A. 1996, \apj\ 459, 558

\markcite{dek86}
	Dekel, A. \& Silk, J. 1986, \apj\ 303, 39

\markcite{deh97}
	Dehnen, W. \& Binney, J. 1997, astro-ph/9612059

\markcite{eng97}
	Englmaier, P., 1997, Ph.D. Thesis, Basel

\markcite{eva94}
	Evans, N. \& Jijina, J. 1994, \mnras\ 267, L21

\markcite{fab83}
	Faber, S. \& Lin, D. 1983, \apj\ 266, L17

\reference{fahl96}
	Fahlman, G., Mandushev, G., Richer, H., Thompson, I., \& Sivaramakrishnan, A.
	1996, \apj\ 459L, 65

\reference{her90}
	Hernquist, L. 1990, \apj\ 356, 359

\reference{me94}
	Ibata, R., Gilmore, G., \& Irwin, M. 1994, \nat\ 370, 194.

\reference{me95c}
	Ibata, R., Gilmore, G., \& Irwin, M. 1995, \mnras\ 277, 781

\reference{me97}
	Ibata, R., Wyse, R., Gilmore, G., Irwin, M. \& Suntzeff, N., 1997,
	\aj\ 113, 634

\reference{irw95}
	Irwin, M. \& \des, D., 1995, \mnras\ 277, 1354

\reference{irw96}
	Irwin, M., Ibata, R., Gilmore, G., Suntzeff, N. \& Wyse, R. 1996,
	in: `Formation of the Galactic Halo', ed~A.~Sarajedini, A.S.P., San Francisco

\reference{joh95}
	Johnston, K. V., Spergel, D. N., \& Hernquist, L. 1995, \apj\ 451, 598

\reference{joh93}
	Johnstone, D. 1993, \aj\ 105, 155

\reference{kin66}
	King, I. 1966, \aj\ 71, 64

\reference{kro97a}
	Kroupa, P., \& Bastian, U. 1997, New Astronomy 2, 77

\reference{kro97b}
	Kroupa, P. 1997,  New Astronomy 2, 139

\reference{lin95}
	Lin, D., Jones, B. \& Klemola, A. 1995, \apj\ 439, 652

\reference{mea88}
	Meatheringham, S., Dopita, M., Ford, H. \& Webster, B. 1988, \apj\
	327, 651

\reference{mat95a}
	Mateo M., Udalski A., Szymanski M., Kaluzny
	J., Kubiak M. \& Kreminski W. 1995a, \aj\ 109, 588

\reference{mat96b}
	Mateo, M., Mirabal, N., Udalski, A., Szymanski, M., Kaluzny,
	J., Kubiak, M., Kreminski, W. \& Stanek, K. 1996, \apj\ 458, L13

\reference{miy75}
	Miyamoto, M. \& Nagai, R. 1975, \pasj\ 27, 533

\reference{oh95}
	Oh, K.S., Lin, D.N. \& Aarseth, S.J. 1995, \apj\ 442, 142

\reference{osw96}
	Oswalt, T., Smith, J., Wood, M. \& Hintzen, P. 1996, \nat\ 382, 692

\reference{pia95}
	Piatek, S. \& Pryor, C. 1995, \aj\ 109, 1071

\reference{pry93}
	Pryor, C. \& Meylan, G. 1993,
	in: `Structure and Dynamics of Globular Clusters',
	eds~S.~Djorgovski and G.~Meylan, A.S.P., San Francisco

\reference{ric93}
	Richardson, D. 1993, Ph.D. Thesis, Cambridge

\reference{ric96}
	Richer, H., Harris, W., Fahlman, G., Bell, R., Bond, H., Hesser, J.,
	Holland, S., Pryor, C., Stetson, P., Vandenberg, D. \& van den
	Bergh, S. 1996, \apj\ 463, 602

\reference{san65}
	Sandage, A., 1965, in `The Structure and Evolution of Galaxies',
	ed.~H.~Bondi (New York, Interscience), p83

\reference{spi87}
	Spitzer, L. 1987, `Dynamical Evolution of Globular Clusters',
	(Princeton University Press, Princeton)

\reference{tot92}
	T\'oth, G. \& Ostriker, J., 1992, \apj\ 389, 5

\reference{una96}
	Unavane, M., Wyse, R. \& Gilmore, G. 1996, \mnras\ 278, 727
	
\reference{vel95}
	Velasquez, H. \& White, S. 1995, \mnras\ 275, L23

\reference{wei95}
	Weinberg, D. 1995, \apj\ 455, L31

\end{references}
\end{document}